\documentclass[apj,numberedappendix]{emulateapj-rtx4}
\usepackage{amssymb,amsmath}
\usepackage{graphicx}
\usepackage{appendix}
\usepackage{natbib}
\usepackage{hyperref}
\usepackage{enumerate}
\usepackage{multirow}
\usepackage{bm}

\newcommand{\etal}{et~al.~}

\altaffiltext{\MIT}{Kavli Institute for Astrophysics and Space Research, Massachusetts Institute of Technology, 77 Massachusetts Avenue, Cambridge, MA 02139}
\altaffiltext{\IfA}{Institute for Astronomy (IFA), University of Hawaii, 2680 Woodlawn Drive, HI 96822}
\altaffiltext{\CfA}{Harvard-Smithsonian Center for Astrophysics, 60 Garden Street, Cambridge, MA 02138}
\altaffiltext{\Harvard}{Department of Physics, Harvard University, 17 Oxford Street, Cambridge, MA 02138}
\altaffiltext{\KIPAC}{Kavli Institute for Particle Astrophysics and Cosmology, Stanford University, 452 Lomita Mall, Stanford, CA 94305}
\altaffiltext{\Stanford}{Department of Physics, Stanford University, 382 Via Pueblo Mall, Stanford, CA 94305}
\altaffiltext{\SLAC}{SLAC National Accelerator Laboratory, 2575 Sand Hill Road, Menlo Park, CA 94025}
\altaffiltext{\AlfA}{Argelander-Institut f{\"u}r Astronomie, Auf dem H{\"u}gel 71, D-53121 Bonn, Germany}
\altaffiltext{\FNAL}{Fermi National Accelerator Laboratory, Batavia, IL 60510-0500}
\altaffiltext{\KICPChicago}{Kavli Institute for Cosmological Physics, University of Chicago, 5640 South Ellis Avenue, Chicago, IL 60637}
\altaffiltext{\AAUChicago}{Department of Astronomy and Astrophysics, University of Chicago, 5640 South Ellis Avenue, Chicago, IL 60637}
\altaffiltext{\ANL}{Argonne National Laboratory, 9700 S. Cass Avenue, Argonne, IL, USA 60439}
\altaffiltext{\UChicago}{Department of Physics, University of Chicago, 5640 South Ellis Avenue, Chicago, IL 60637}
\altaffiltext{\Miss}{Department of Physics and Astronomy, University of Missouri, 5110 Rockhill Road, Kansas City, MO 64110}
\altaffiltext{\EFIChicago}{Enrico Fermi Institute, University of Chicago, 5640 South Ellis Avenue, Chicago, IL 60637}
\altaffiltext{\Munich}{Department of Physics, Ludwig-Maximilians-Universit\"{a}t, Scheinerstr.\ 1, 81679 M\"{u}nchen, Germany}
\altaffiltext{\ExcellenceCluster}{Excellence Cluster Universe, Boltzmannstr.\ 2, 85748 Garching, Germany}
\altaffiltext{\UFlorida}{Department of Astronomy, University of Florida, Gainesville, FL 32611}
\altaffiltext{\UMontreal}{D\'{e}partement de Physique, Universit\'{e} de Montr\'{e}al, C.P. 6128, Succ. Centre-Ville, Montr\'{e}al, Qu\'{e}bec H3C 3J7, Canada}
\altaffiltext{\Berkeley}{Department of Physics, University of California, Berkeley, CA 94720}
\altaffiltext{\Arizona}{Steward Observatory, University of Arizona, 933 North Cherry Avenue, Tucson, AZ 85721}
\altaffiltext{\Melbourne}{School of Physics, University of Melbourne, Parkville, VIC 3010, Australia}
\altaffiltext{\CaseWestern}{Physics Department, Center for Education and Research in Cosmology and Astrophysics, Case Western Reserve University, Cleveland, OH 44106}
\altaffiltext{\Davis}{Department of Physics, University of California, One Shields Avenue, Davis, CA 95616}
\altaffiltext{\Illinois}{Department of Astronomy and Department of Physics, University of Illinois, 1002 West Green St., Urbana, IL 61801, USA}
\altaffiltext{\CTIO}{Cerro Tololo Inter-American Observatory, La Serena, Chile}

\def\MIT{1}
\def\IfA{2}
\def\CfA{3}
\def\Harvard{4}
\def\KIPAC{5}
\def\Stanford{6}
\def\SLAC{7}
\def\AlfA{8}

\def\FNAL{9}
\def\KICPChicago{10}
\def\AAUChicago{11}
\def\ANL{12}
\def\UChicago{13}
\def\Miss{14}
\def\EFIChicago{15}
\def\Munich{16}
\def\ExcellenceCluster{17}
\def\UFlorida{18}
\def\UMontreal{19}
\def\Berkeley{20}
\def\Arizona{21}
\def\Melbourne{22}
\def\CaseWestern{23}
\def\Davis{24}
\def\Illinois{25}
\def\CTIO{26}

\begin{document}

%% LaTeX will automatically break titles if they run longer than
%% one line. However, you may use \\ to force a line break if
%% you desire.

\title{Star-Forming Brightest Cluster Galaxies at $0.25 < \lowercase{z} < 1.25$: A Transitioning Fuel Supply}

%% Use \author, \affil, and the \and command to format
%% author and affiliation information.
%% Note that \email has replaced the old \authoremail command
%% from AASTeX v4.0. You can use \email to mark an email address
%% anywhere in the paper, not just in the front matter.
%% As in the title, use \\ to force line breaks.

%\author{The SPT Team}
%auto-ignore

\author{
M.~McDonald\altaffilmark{\MIT},
B.~Stalder\altaffilmark{\IfA},
M.~Bayliss\altaffilmark{\CfA,\Harvard},
S.~W.~Allen\altaffilmark{\KIPAC,\Stanford,\SLAC},
D.~E.~Applegate\altaffilmark{\AlfA},
M.~L.~N.~Ashby\altaffilmark{\CfA},
M.~Bautz\altaffilmark{\MIT},
B.~A.~Benson\altaffilmark{\FNAL,\KICPChicago,\AAUChicago},
L.~E.~Bleem\altaffilmark{\KICPChicago,\ANL,\UChicago},
M.~Brodwin\altaffilmark{\Miss},
J.~E.~Carlstrom\altaffilmark{\KICPChicago,\AAUChicago,\ANL,\UChicago,\EFIChicago}, 
I.~Chiu\altaffilmark{\Munich},
S.~Desai\altaffilmark{\Munich,\ExcellenceCluster},
A.~H.~Gonzalez\altaffilmark{\UFlorida},
J.~Hlavacek-Larrondo\altaffilmark{\UMontreal},
W.~L.~Holzapfel\altaffilmark{\Berkeley},
D.~P.~Marrone\altaffilmark{\Arizona},
E.~D.~Miller\altaffilmark{\MIT},
C.~L.~Reichardt\altaffilmark{\Melbourne},
B.~R.~Saliwanchik\altaffilmark{\CaseWestern}, 
A.~Saro\altaffilmark{\Munich},
T.~Schrabback\altaffilmark{\AlfA},
S.~A.~Stanford\altaffilmark{\Davis}
A.~A.~Stark\altaffilmark{\CfA}, 
J.~D.~Vieira\altaffilmark{\Illinois},
A.~Zenteno\altaffilmark{\CTIO}
}

\email{mcdonald@space.mit.edu}   %optional

%% Mark off your abstract in the ``abstract'' environment. In the manuscript
%% style, abstract will output a Received/Accepted line after the
%% title and affiliation information. No date will appear since the author
%% does not have this information. The dates will be filled in by the
%% editorial office after submission.

\begin{abstract}
We present a multi-wavelength study of 90 brightest cluster galaxies (BCGs) in a sample of galaxy clusters selected via the Sunyaev Zel'dovich effect by the South Pole Telescope, utilizing data from various ground- and space-based facilities. We infer the star formation rate (SFR) for the BCG in each cluster, based on the UV and IR continuum luminosity, as well as the [O\,\textsc{ii}]$\lambda\lambda$3726,3729 emission line luminosity in cases where spectroscopy is available, finding 7 systems with SFR $>$ 100 M$_{\odot}$ yr$^{-1}$. We find that the BCG SFR exceeds 10 M$_{\odot}$ yr$^{-1}$ in 31 of 90 (34\%) cases at $0.25 < z < 1.25$, compared to $\sim$1--5\% at $z\sim0$ from the literature. At $z\gtrsim1$, this fraction increases to 92$^{+6}_{-31}$\%, implying a steady decrease in the BCG SFR over the past $\sim$9 Gyr. At low-$z$, we find that the specific star formation rate in BCGs is declining more slowly with time than for field or cluster galaxies, most likely due to the replenishing fuel from the cooling ICM in relaxed, cool core clusters. 
At $z\gtrsim0.6$, the correlation between cluster central entropy and BCG star formation -- which is well established at $z\sim0$ -- is not present. Instead, we find that the most star-forming BCGs at high-$z$ are found in the cores of dynamically unrelaxed clusters. We investigate the rest-frame near-UV morphology of a subsample of the most star-forming BCGs using data from the \emph{Hubble Space Telescope}, finding complex, highly asymmetric UV morphologies on scales as large as $\sim$50--60 kpc.  The high fraction of star-forming BCGs hosted in unrelaxed, non-cool core clusters at early times suggests that the dominant mode of fueling star formation in BCGs may have recently transitioned from galaxy-galaxy interactions to ICM cooling.
%The apparent correlation between the dynamical state of the cluster and the BCG SFR at early times suggests that the primary mode of star formation may transition from merger-driven at $z\gtrsim0.7$ to cooling-induced at $z\lesssim0.7$.

%% Keywords should appear after the \end{abstract} command. The uncommented
%% example has been keyed in ApJ style. See the instructions to authors
%% for the journal to which you are submitting your paper to determine
%% what keyword punctuation is appropriate.
\end{abstract}

% \vspace{-0.2in}}

%================================================================%
%============== INTRODUCTION ====================================%
%================================================================%
\section{Introduction}
\setcounter{footnote}{0}

One of the great mysteries in astronomy today is why $\gtrsim$90\% of the baryons in the Universe, which are in diffuse gas with relatively short cooling times \citep[e.g.,][]{shull12}, have not cooled and formed stars. This inefficient star formation manifests as a significant disagreement between the predicted galaxy luminosity function from $\Lambda$CDM cosmological simulations and that observed in the local Universe. In the most massive galaxies in the Universe, which are found at the centers of rich galaxy clusters, this disagreement is maximized, with central cluster galaxies being substantially less massive than predicted by simple models \citep[see review by][]{silk12}. This has become known as the ``cooling flow problem'' and can be stated simply as: ``Why, given the short cooling time of the intracluster medium in the cores of some galaxy clusters, do we not observe  massive starburst galaxies at the centers of these clusters?''.

Over the past couple of decades, much effort has been devoted to answering this question. Early work focused on searching for multiphase gas and star formation in brightest cluster galaxies (BCGs). Numerous studies have found evidence for ultraviolet (UV) and infrared (IR) continuum \citep[e.g.,][]{mcnamara89, hicks05,odea08,mcdonald11b,hoffer12, fraser14,donahue15}, warm, ionized gas \citep[e.g.,][]{hu85,johnstone87,heckman89,crawford99, edwards07, hatch07, mcdonald10, mcdonald11a}, and both warm and cold molecular gas \citep[e.g.,][]{jaffe97, donahue00, edge01, edge02, edge03, salome03,hatch05, jaffe05, johnstone07, oonk10,mcdonald12b} -- all of which are indicative of ongoing or recent star formation. Star-forming BCGs were found preferentially in galaxy clusters with ``cool cores'', as identified by a central density enhancement in the ICM \citep[e.g.,][]{vikhlinin07,santos08,hudson10} or low central entropy/cooling time \citep[e.g.,][]{cavagnolo08,cavagnolo09,hudson10}. These and other works established a link between the cooling ICM and the presence of multiphase gas, suggesting that cooling flows may indeed be fueling star formation in BCGs. However, the typical star formation rates inferred from a variety of indicators were found to be only $\sim$1\% of the expected ICM cooling rate \citep[e.g.,][]{odea08}. Roughly one of the two orders of magnitude in this disagreement can be accounted for by inefficient star formation \citep{mcdonald14b}, however a further order-of-magnitude disagreement between the cooling predictions and observations still remains.

Some form of feedback is necessary to prevent the bulk ($\sim$90\%) of the cooling ICM from becoming fuel for star formation. The leading candidate is ``radio-mode'' feedback \citep[see reviews by][]{fabian12,mcnamara12} from active galactic nuclei (AGN), which are ubiquitous at the centers of cool core clusters \citep{sun09b}. The mechanical energy output from these AGN are sufficient to offset cooling on large scales, preventing runaway cooling in the majority of clusters \citep[e.g.,][]{birzan04, rafferty08, hlavacek12,hlavacek14} with a few notable exceptions \citep{mcnamara06,mcdonald12c}. The low levels of star formation and gas in multiphase filaments are understood to be local thermodynamic instabilities \citep[e.g.,][]{sharma10,gaspari12,mccourt12,voit15a} in regions where, \emph{locally}, cooling dominates over feedback, despite the overall global balance. 
These star formation rates, which average a few M$_{\odot}$ yr$^{-1}$, may contribute a few percent to the total stellar mass of the BCG over the past $\sim$8 Gyr -- the majority of the growth in these systems likely comes from ``dry mergers'' (mergers of gas-poor galaxies), which increase the stellar mass by a factor of $\sim$2 from $z=1$ to $z=0$ \citep{ruszkowski09}.

The picture presented here is based almost entirely on observations of nearby ($z\lesssim0.3$) galaxy clusters. This is, in part, due to the fact that these clusters are more easily studied because of their proximity (improved signal-to-noise, angular resolution, etc). Equally important, however, is the scarcity of well-understood samples of high-redshift galaxy clusters. Until recent years there were few samples of galaxy clusters with known masses at $z\gg0.5$ -- surveys that did probe high redshift clusters were generally flux-limited or were assembled from serendipitous detections via a heterogeneous collection of methods. With the advent of large-area mm-wave surveys utilizing the Sunyaev Zel'dovich (SZ) effect \citep{sunyaev72} to detect galaxy clusters, this situation has changed dramatically over the past several years, with the latest surveys achieving nearly redshift-independent detection of clusters above a fixed mass threshold at $z\gtrsim0.3$. Most recently, the completed 2500 deg$^2$ SZ survey with the South Pole Telescope \citep[SPT;][]{carlstrom11} has discovered more than 500 massive galaxy clusters, the majority of which are at $z>0.5$ \citep{bleem15}. In this work, we focus on a subsample of this survey which has been targeted for X-ray follow-up \citep{mcdonald13b} and spans a redshift range of $0.3 < z < 1.2$. The availability of optical photometry and spectroscopy for the majority of the BCGs in this sample, along with archival UV (GALEX) and IR (WISE) data allows us to study star formation in BCGs at high redshift for a complete, mass-selected sample. 

The remainder of this paper is structured as follows. In \S2 we define the sample used in this work, and present the multiwavelength data and analysis techniques that will enable us to identify star-forming BCGs. In \S3 we isolate the sample of star-forming BCGs and attempt to determine whether there is any evolution in their properties or the properties of their host clusters. In \S4 we discuss these results, addressing bias and selection concerns, while trying to draw a broad picture of galaxy and galaxy cluster evolution within which these results fit. We finish in \S5 with a brief summary of the important results of this work, and a look toward the future. Throughout this work we assume H$_0$ = 70 km s$^{-1}$ Mpc$^{-1}$, $\Omega_M$ = 0.27, $\Omega_{\Lambda}$ = 0.73, and a \cite{salpeter55} initial mass function (IMF).

\section{Data \& Analysis}
\subsection{Cluster Sample and BCG Selection}
We initially define the sample to include all 83 clusters from \cite{mcdonald13b}, which were selected from the SPT 2500 deg$^2$ survey \citep{bleem15} and subsequently observed in the X-ray with the \emph{Chandra X-ray Observatory}. To this sample, we add an additional 8 clusters from \cite{bleem15} that have archival \emph{Chandra} data from other sources (SPT-CLJ0106-5943, SPT-CLJ0232-4421, SPT-CLJ0235-5121, SPT-CLJ0516-5430, SPT-CLJ0522-4818, SPT-CLJ0658-5556, SPT-CLJ2011-5725, SPT-CLJ2332-5053). X-ray data products, such as central entropy (K$_0$), are presented for this full sample of 91 clusters in \cite{mcdonald13b} -- we direct the reader there for a full description of our X-ray methodology.

For each cluster, we have obtained some combination of ground-based $g,r,i,z$ optical imaging, ground-based J, H, K near-IR imaging, and \emph{Spitzer} IRAC 3.6$\mu$m, 4.5$\mu$m imaging, as described in detail in \cite{bleem15}. These data have been used to provide optical confirmation of the SZ-selected clusters, and to estimate a photometric redshift based on the member galaxy colors. We remove from our sample all clusters for which the optical-IR follow-up was performed in $\leq$2 filters, preventing a reliable fit to the stellar continuum. After this cut, we were left with a sample of 82 clusters. From this follow-up imaging, we select the BCG\footnote{We note here that the term ``BCG'' is largely inappropriate for this work, since there may be brighter galaxies within the virial radius (e.g., AGN). The more appropriate monicker would be ``central cluster galaxy'', but we opt for BCG throughout this paper since it is more commonly used.}  initially as the brightest red-sequence galaxy within an aperture of R$_{200}$ centered on the SZ peak, following \cite{song12}. We then visually inspect each cluster and select a new BCG if one, or both, of two conditions are met: i) there is a similarly-bright galaxy that is significantly closer to the X-ray peak (29\% of systems); ii) there is a bright blue galaxy on or near the X-ray peak that was not selected due to our initial preference for red galaxies (3\% of systems). Wherever possible, we use HST imaging from the SPT weak-lensing follow-up programs with IDs 12246, 12477, 13412 (PIs: Stubbs, High, Schrabback) to aid in the visual identification of the BCG.

%
%\begin{itemize}
%\item There is a similarly-bright galaxy that is significantly closer to the X-ray peak (26/91).
%\vspace{ -0.03in}
%\item There is a bright blue galaxy on or near the X-ray peak that was not selected due to our initial preference for red galaxies (3/91).
%\vspace{ -0.03in}
%\item There is a blue point source coincident with the X-ray peak (possible QSO-BCG) (3/91).
%\end{itemize}

In all cases where it was not clear which galaxy was the BCG, we retain the original red BCG, allowing for two BCGs in these clusters. In this case, we give each BCG a 50\% weight when performing statistical analysis and include the limiting cases in our error estimates (i.e., upper limit contains star-forming BCG, lower limit contains passive BCG). This procedure resulted in a sample of 90 BCGs in 82 clusters. The net effect of allowing multiple BCGs in cases where identification of a single object was challenging is to increase our uncertainties, increasing the likelihood that the ``true'' answer lies without our error bars.

\subsection{X-ray Analysis: Central Entropy and Luminosity}
Several studies have found correlations between the amount of star formation in the BCG and the core entropy \citep[e.g.,][]{cavagnolo08,odea08,voit15b} and luminosity-derived cooling rate \citep[e.g.,][]{crawford99,odea08,mcdonald10,mcdonald11b}. This has led to the conclusion that star formation is being fueled in the BCG by residual cooling flows \citep{voigt04,tremblay12,mcdonald14b}. In order to test whether this trend was established at high-$z$, we require estimates of the core entropy and cooling rate for each cluster. Given that we only have $\sim$2000 X-ray counts per cluster, modeling the central entropy \citep[e.g.,][]{cavagnolo09} or estimating the spectroscopically-derived cooling rate \citep[e.g.,][]{voigt04} is not feasible. Instead, we compute spectroscopic quantities (bolometric luminosity, temperature) from a circular aperture with radius of 0.075\,R$_{500}$ \citep[where R$_{500}$ was derived based on the Y$_X$--M$_{500}$ relation of][]{vikhlinin09a}, which should roughly correspond to the deprojected core temperature \citep[see e.g.,][]{mcdonald14c}. X-ray spectra extracted from this aperture are modeled with a photometric absorption (\textsc{phabs}) and plasma (\textsc{apec}) model, allowing the temperature, metallicity, and normalization of the plasma model to vary. This choice of aperture is meant to capture the ``core'' properties, reflecting the realistic fuel reservoir that the BCG may have access to. 

The ``central'' density is computed based on the deprojected X-ray surface brightness profile, following \cite{vikhlinin06a} and \cite{mcdonald13b}. Since the measurement of electron density requires far fewer X-ray counts than the measurement of spectroscopic temperature, the central density is measured at $r=0.01$R$_{500}$, or roughly 10 kpc for a typical cluster in this sample. We combine the projected core temperature and deprojected core density to arrive at a pseudo-deprojected core entropy ($K_0 = kT_{0.075R_{500}}\,n_{e,0.01R_{500}}^{-2/3}$).

Unlike our previous work \citep{mcdonald13b}, here we measure ``central'' quantities at the X-ray peak, rather than the large-scale centroid. In general, the BCG is located closer to the X-ray peak than the centroid in cases of merging clusters, which motivated this choice. We will discuss in later sections the effects of this choice. For a more detailed description of our X-ray analysis techniques, we direct the reader to \cite{mcdonald13b} and \cite{mcdonald14c}.

\subsection{UV--Optical--IR Photometry and SED Modeling}
Ground-based optical and/or near-infrared imaging for all clusters in this sample have been obtained at a variety of wavelengths as part of a confirmation and photometric redshift follow-up campaign. The acquisition, reduction, and calibration of these data are presented in detail in \cite{song12} and \cite{bleem15}. Aperture photometry for the BCG was obtained from SExtractor \citep{bertin96}, following \cite{song12} and \cite{bleem15}.

In addition to these existing data, we have acquired new ground-based $u$-band imaging with Megacam \citep{megacam} on the Magellan Clay telescope for 49 clusters in this sample. These 49 clusters were those that had the least restrictive upper limits on the BCG SFR at other wavelengths, generally lacking in spectroscopic or deep GALEX coverage. The exposure time for these observations was chosen to provide an overall sensitivity of our survey to obscured star formation rates of $\sim$10 M$_{\odot}$ yr$^{-1}$ -- without these additional data, our sensitivity limit would vary significantly with redshift. These data were reduced using the standard photometric pipeline described in \cite{bleem15}.

The position of each BCG was cross-referenced with the \emph{GALEX}\footnote{http://galex.stsci.edu/GR6/} \citep{galex} and \emph{WISE}\footnote{http://irsa.ipac.caltech.edu/Missions/wise.html} \citep{wise} archives, assuming a maximum offset of 2$^{\prime\prime}$, from which we obtained near-UV (NUV) and near--mid IR (NIR, MIR) photometry for each BCG. If the BCG was undetected by either of these surveys, we instead obtained upper limits. \emph{Spitzer} 3.6$\mu$m, 4.5$\mu$m, and \emph{WISE} 4-band photometry was converted from Vega to AB magnitudes following \cite{price04}.
The resulting UV--optical--IR spectral energy distributions (SEDs), which span 2000\AA\ to 22$\mu$m in the observed frame, are shown in Figure \ref{fig:seds} for our sample of 90 BCGs. As mentioned above, we have discarded all BCGs for which we have $\le$2 photometric measurements (excluding upper limits).

The observed SEDs are fit in two stages. First, we model the full SED with a single-age population with formation redshift $z_f$, solar metallicity, and a Salpeter \citep{salpeter55} initial mass function. We assume a uniformly-distributed range of $z_f$ from 2--5 in our models, which leads to some uncertainty in our resulting stellar masses and star formation rates.
The spectrum for this old stellar population was generated using \textsc{Starburst99} \citep{leitherer99} and was convolved with our broadband filter set. The model spectrum was fit to the data using \textsc{mpfitfun}\footnote{http://www.physics.wisc.edu/$\sim$craigm/idl/down/mpfitfun.pro}, which minimizes $\chi^2$ with respect to the two free parameters (normalization, redshift). The redshift was allowed to vary within the measured uncertainty from \cite{bleem15}, while the normalization, which corresponds to the stellar mass of the old population, was left free.

Beyond this single-component model, we also consider additional components in the UV and IR representing contributions from a young stellar population and warm dust, respectively. At short wavelengths, we model excess emission using \textsc{Starburst99}, assuming a constant star formation rate over the past 30\,Myr (roughly the AGN duty cycle). We note that adjusting this timescale down to 10\,Myr or up to 100\,Myr results in deviations in the derived SFR of $\sim$20\%. We assume that the emission from young stars is obscured by dust, incorporating the \cite{calzetti00} extinction law and an intrinsic reddening of $E(B-V) = 0.3\pm0.1$ -- this range is based on observations of nearby star-forming BCGs \citep{crawford99,mcdonald12a}. The young stellar component has a single free parameter, corresponding to the ongoing, extinction-corrected star formation rate. At long wavelengths, we mimic a dusty component with a mid-infrared power-law ($F_{\nu} \propto \lambda^{2.0\pm0.5}$), following \cite{casey12}. This dust component is artificially truncated at shorter wavelengths, so that it will not add UV flux. We only include the young and dusty components if their inclusion improves the $\chi^2_{dof}$, which is the case for 34\% (31/90) of the BCGs in our sample. For the remaining 66\% (59/90), the ``passive-evolution'' model yields a suitable fit to the data. The results of this SED-fitting are shown in Figure \ref{fig:seds}.

\subsection{Optical Spectroscopy}
For 36 of the 95 BCGs in this sample, we have optical spectroscopy from a combination of the IMACS \citep{imacs} and LDSS3 spectrographs on Magellan. These spectra were initially obtained as part of a spectroscopic redshift campaign which targeted, primarily, red sequence galaxies. The full details of this spectroscopic follow-up is provided in \cite{ruel13} and \cite{bleem15}. For each spectrum, we measure the [O\,\textsc{ii}] and H$\delta$ equivalent widths and the 4000\AA\ break strength (D$_{4000}$), via indices defined by \cite{balogh99}. [O\,\textsc{ii}] equivalent widths are converted to emission line fluxes using the continuum level, as determined by the SED modeling described in \S2.2, at rest-frame 3727\AA.
From these 36 spectra we find significant ($>$3$\sigma$) [O\,\textsc{ii}] emission in 5 systems, one of which is the Phoenix cluster \citep{mcdonald12c}. This system allows us an opportunity to compare the index-based measurement of [O\,\textsc{ii}] to the total, flux from deep integral-field spectroscopy \citep{mcdonald14a}. Using the long-slit data, the index-based technique of \cite{balogh99} combined with our best-fit optical SED yields a line flux of $f_{[O II]} = 2.2 \times 10^{-14}$ ergs s$^{-1}$ cm$^{-2}$ for the central galaxy in the Phoenix cluster, while the total measured flux from \cite{mcdonald13a} is $f_{[O II]} = 1.6 \times 10^{-14}$ ergs s$^{-1}$ cm$^{-2}$. Thus, for this single comparison, the index-based approach does an adequate job of reproducing the measurement made via emission-line modeling of significantly higher-quality data. 

When correcting the [O\,\textsc{ii}] emission line flux for extinction, we assume that the ionized gas has a factor of two higher reddening than the stellar continuum, following \cite{calzetti00,calzetti01}. This assumption is motivated by the idea that the warm ionized gas traces the highly-ionizing O and B stars, which tend to be embedded in dusty, star-forming regions.

\subsection{Star Formation Rates and Their Uncertainties}

\subsubsection{UV-Derived SFRs}
The calculation of star formation rates (SFRs) from the UV photometry is straightforward. In \S2.3 we describe our SED-fitting procedure, which includes an intrinsically-reddened, continuously-star forming population derived from \textsc{Starburst99} models. The normalization of this component yields the current star formation rate, under the assumption of constant star formation for the past 30~Myr.

The uncertainty in the UV-derived SFR is dominated by our uncertainty in the amount of intrinsic extinction ($E(B-V)$) and, to a lesser extent, the formation redshift of the old stellar population ($z_f$). In order to propagate our uncertainty in these quantities to our estimate of the SFR we perform 100 fits to each SED, varying $z_f$ and $E(B-V)$ in each fit. We assume a normal distribution for $E(B-V)$, with $E(B-V) = 0.3 \pm 0.1$, motivated by observations of nearby star-forming BCGs \citep{crawford99,mcdonald12a}. For the old stellar population, we assume a uniform distribution of formation epochs, from $z_f = 2.0 - 5.0$ Gyr. This Monte-Carlo approach results in typical uncertainties in the UV-derived SFRs of 0.38 dex, or a factor of $\sim$2.5.

\subsubsection{[O\,\textsc{ii}]-Derived SFRs}
As described in \S2.3, [O\,\textsc{ii}] fluxes are derived from a combination of indices \citep{balogh99}, which provide an estimate of the equivalent line width, and SED fitting, which provides the interpolated continuum level at 3727\AA. We convert the measured emission line flux to a SFR, assuming 
SFR$_{[OII]} =  9.53 \times 10^{-42}$ L$_{[O II]}$ \citep{kewley04}. 
As in \S2.5.1, we assume a normal distribution of intrinsic reddening, with $E(B-V)_{gas} = 2E(B-V)_{stars}$, following \cite{calzetti00,calzetti01}. As is the case for the UV-derived SFRs, our uncertainty in this extinction corrections dominates the uncertainty in the inferred SFR. Propagating this uncertainty through the fitting procedure via a Monte Carlo approach yields typical uncertainties of 0.19 dex, or a factor of $\sim$1.5.

\subsubsection{IR-Derived SFRs}
In the mid-infrared, we derive the SFR by first extrapolating our power-law fit from observed 22$\mu$m to rest-frame 24$\mu$m. The uncertainty on this calculation is both redshift dependent, since at higher redshift we are extrapolating over a larger wavelength range, and dependent on the assumed power-law slope. We incorporate the uncertainty in the power-law slope \citep[$\alpha = 2.0 \pm 0.5$;][]{casey12} in our calculation of the 24$\mu$m luminosity by assuming a normal distribution of values and performing 100 fits to the data. The average resulting uncertainty in L$_{24\mu m}$ is 0.2 dex, or a factor of $\sim$1.6. These extrapolated values of L$_{24\mu m}$ are then converted into an estimate of the SFR, following \cite{calzetti07}.

\subsubsection{Comparison of SFR Estimates}
\begin{figure}[b]
\centering
\includegraphics[width=0.48\textwidth]{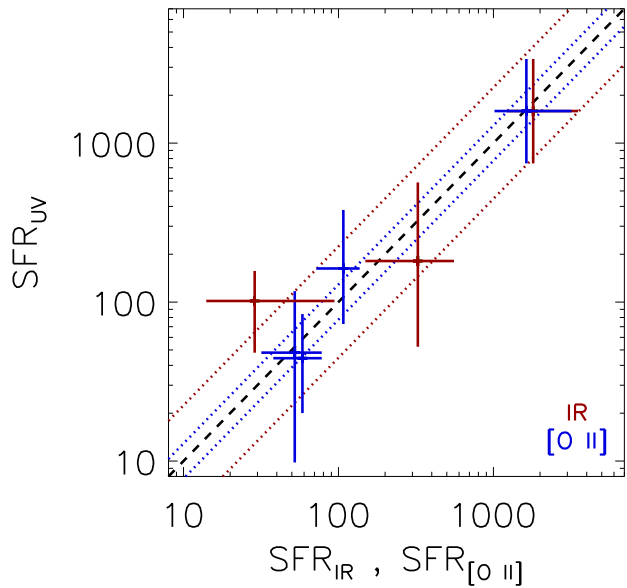}
\caption{Comparison of SFRs derived via UV, [O\,\textsc{ii}], and IR relations for all 6 systems with multiple measurements. The dashed line represents the one-to-one relation, while the dotted lines show the measured scatter between the UV--IR (red) and UV--[O\,\textsc{ii}] (blue) relations. The dominant source of uncertainty for UV- and [O\,\textsc{ii}]-derived SFRs is the extinction correction, while for IR it is the extrapolation from observed-frame 22$\mu$m to rest-frame 24$\mu$m. This figure demonstrates the overall agreement between these different indicators, despite the significant uncertainties involved in estimating the SFR.}
\label{fig:sfrs}
\end{figure}

In Figure \ref{fig:sfrs}, we compare SFRs derived from the three indicators described above. We have only limited overlap between the three subsamples, with only 3 BCGs having both IR- and UV-derived SFRs and an additional 4 BCGs with both UV- and [O\,\textsc{ii}]-derived SFRs. Only a single BCG (SPT-CLJ2344-4243) has SFRs inferred from all three methods. Nonetheless, we proceed to compare how well these estimates agree for systems in common. In general, SFR estimates from different indicators agree within the systematic errors. 
Assuming a one-to-one relation, we measure a scatter of 0.37 dex between UV- and IR-derived SFRs, and 0.15 dex between UV- and [O\,\textsc{ii}]-derived SFRs. The better agreement between the UV- and [O\,\textsc{ii}]-derived SFRs is most likely due to the fact that the dominant systematic uncertainties (extinction, old stellar population age) are correlated between these SFR estimates. 
SFRs for each BCG are listed in Table \ref{table:sample}, along with relevant information about the host cluster.

\subsection{AGN Contamination}
Before presenting results from this survey, we would like to draw attention to one potential issue with our approach. All three indicators of star formation employed here -- UV continuum, [O\,\textsc{ii}] line emission, and IR continuum -- are also indicators of active nuclei. The relative amounts of contamination at each of these wavelengths depends on the type of AGN considered. For example, in the Phoenix cluster, more than half of the total IR continuum comes from a dusty QSO. On the other hand, this AGN contributes $<$5\% to the UV and [O\,\textsc{ii}] emission. Several of the SEDs shown in Figure \ref{fig:seds} exhibit a powerlaw shape that is consistent with both a dusty starburst and an AGN.

\begin{figure}[htb]
\centering
\includegraphics[width=0.48\textwidth]{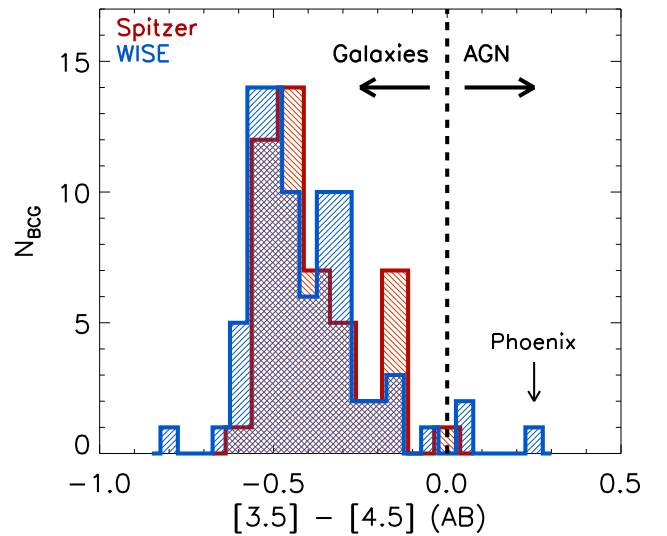} 
\caption{Distribution of mid-IR color, $[3.5]-[4.5]$, for the BCGs in this study from the \emph{Spitzer} (red) and \emph{WISE} (blue) space telescopes. The vertical cut at $[3.5]-[4.5] = 0$, which separates galaxies from AGN, is motivated by \cite{stern05}, and is valid for systems at $z\lesssim1.3$. This figure demonstrates that, with the exception of the Phoenix cluster (SPT-CLJ2344-4243), this sample of BCGs is relatively free of strong AGN.}
\label{fig:spitzer_agn}
\end{figure}

In an attempt to quantify the contamination due to AGN in this sample, we consider in Figure \ref{fig:spitzer_agn} the 3.5$\mu$m -- 4.5$\mu$m color for 49 BCGs with \emph{Spitzer} data and 82 BCGs with \emph{WISE} data. Color corrections from \cite{stern12} have been applied to match data from these telescopes to a common photometric system. Following \cite{stern05,stern12}, we classify systems with mid-IR colors $[3.5]-[4.5] > 0$ as AGN, while those with redder colors are either passive or star-forming galaxies. In this color space, the Phoenix cluster \citep{mcdonald12c} is the only BCG harboring a strong AGN.  At most, we estimate that $\sim$4 clusters in this sample may harbor strong AGN at their centers, based on this mid-IR color selection. This lack of AGN contamination is further confirmed by visual inspection of the \emph{Chandra} data for each cluster, which show a general lack of strong X-ray point sources coincident with the BCG in all clusters with the exception of Phoenix.

We proceed with this work assuming that all systems identified as star-forming are, indeed, star-forming, but remain cognizant of the fact that some fraction of these may host an AGN. We will return to this dilemma in the discussion section.

\begin{figure*}[htb]
\centering
\begin{tabular}{c c}
\includegraphics[width=0.495\textwidth]{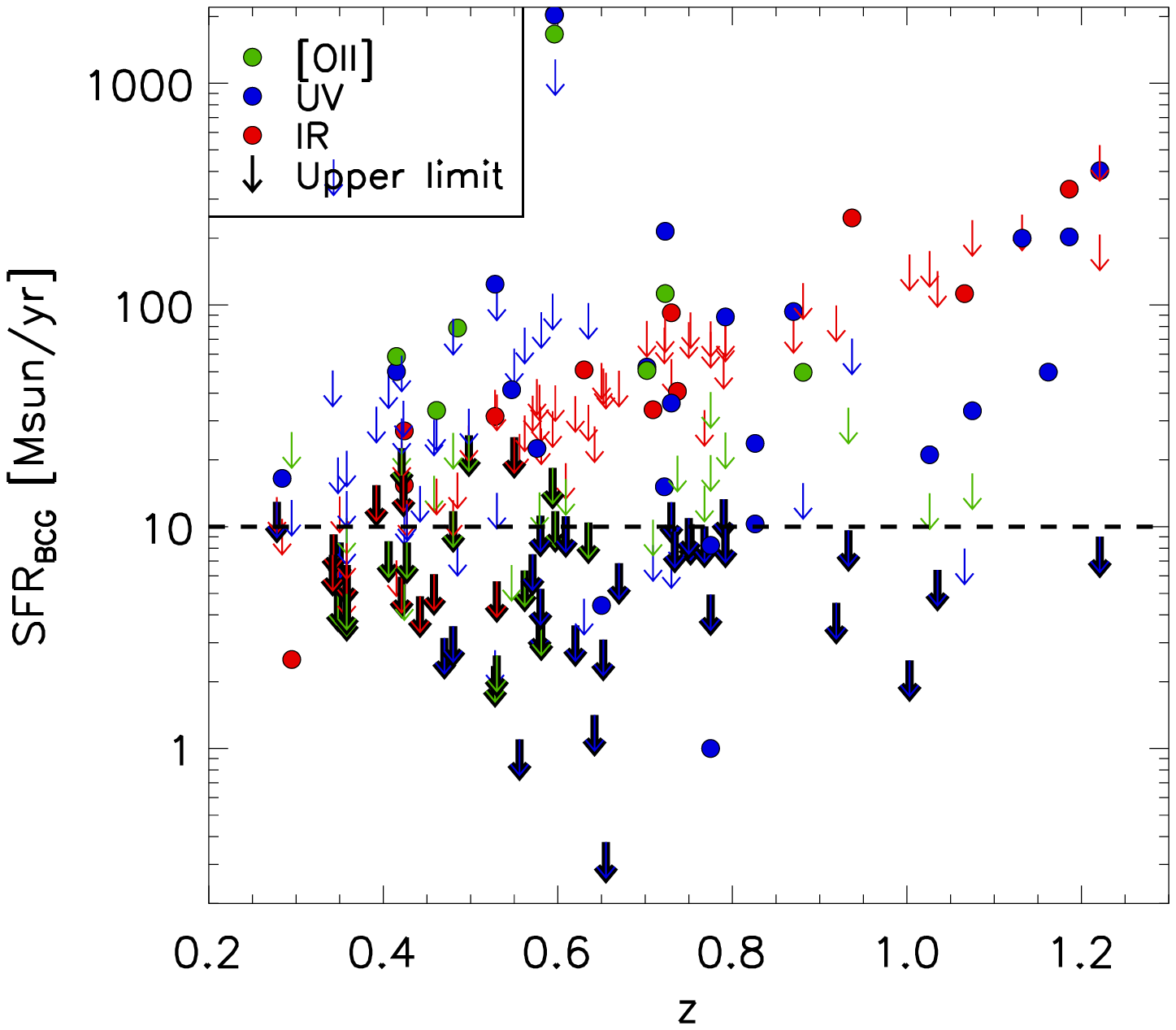} &
\includegraphics[width=0.495\textwidth]{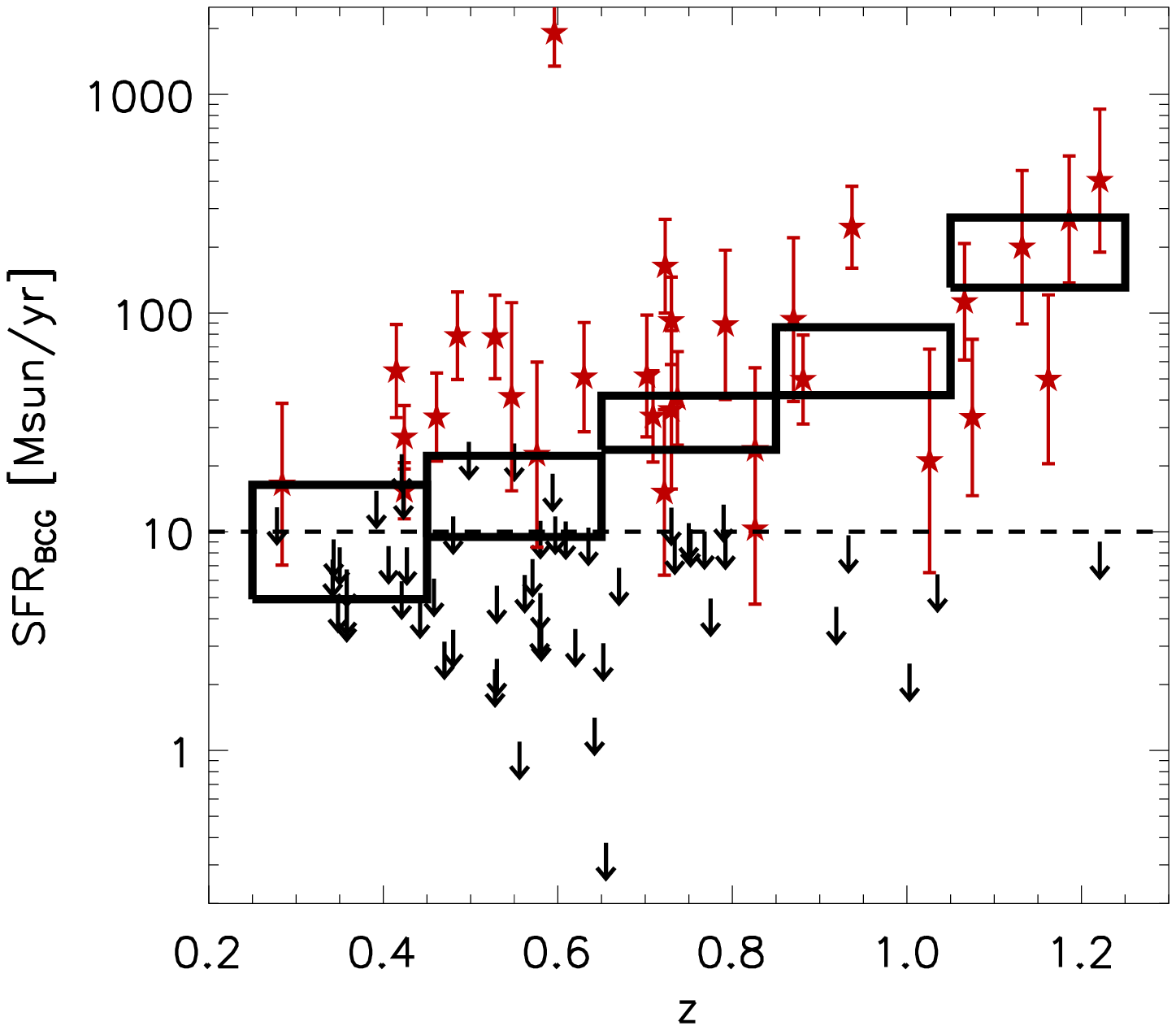} \\
\end{tabular}
\caption{\emph{Left}: Measured star formation rates (filled circles) and upper limits (arrows) for the BCGs in this sample as a function of redshift. We show individual measurements from each method (UV continuum, IR continuum, [O\,\textsc{ii}] emission line) where available. The horizontal dashed line represents our cutoff depth of SFR $= 10$ M$_{\odot}$ yr$^{-1}$. We highlight the most restrictive upper limit for each BCG with a black outline, showing that the best constraints at high-$z$ come from deep rest-frame UV imaging, while at low-$z$ archival data from WISE is able to provide robust upper limits. \emph{Right}: Similar to left panel, but now showing only the most constraining upper limit or detection (red) for each BCG. Where multiple SFRs were measured for a given cluster (see Figure \ref{fig:sfrs}), we take the average. Systematic uncertainties (see \S2.5) are shown as vertical error bars. For UV- and [O\,\textsc{ii}]-based SFRs, the uncertainty is dominated by the extinction correction, while for IR-based SFRs the uncertainty is dominated by the extrapolation from observed 22$\mu$m to rest-frame 24$\mu$m. Black rectangles show the average SFR in different redshift bins, where the height represents the combined statistical uncertainty in the mean and additional uncertainty due to non-detections.}
\label{fig:sfr_z}
\end{figure*}

\section{Results}

Below, we summarize the key results to emerge from this data set. We defer a detailed discussion of these results to the discussion section.

\subsection{Star Formation Rates in BCGs at 0.25 $<$ z $<$ 1.2}

In Figure \ref{fig:sfr_z} we show the star formation rates, derived via three different methods, for the sample of BCGs described in \S2.1. This plot demonstrates that the WISE mid-IR data typically provide the best upper limits at low-redshift, but are unable to provide meaningful upper limits on the star formation rate for BCGs at $z\gtrsim0.5$. At $z>0.5$, the most restrictive limit generally comes from our deep $u$-band follow-up program, which was designed to achieve a redshift-independent sensitivity. For the most part, we are sensitive to SFRs higher than 10 M$_{\odot}$ yr$^{-1}$, with $<$10\% of systems having limits higher than this threshold. 

In the right panel of Figure \ref{fig:sfr_z}, we combine the constraints from the three different SF indicators. This plot shows a significant number of BCGs with SFRs from 10-300 M$_{\odot}$ yr$^{-1}$ at $z>0.4$. One quarter (7/31) of the star-forming BCGs have SFR $>$ 100 M$_{\odot}$ yr$^{-1}$ -- of these, 5 are at $z>0.9$. For comparison, prior to this work there were only four confirmed clusters with extinction-corrected SFR $>$ 100 M$_{\odot}$ yr$^{-1}$ in their BCG: Abell 1835 \citep{mcnamara06}, RX~J1504.1-0248 \citep{ogrean10}, MACS~J1931.8-2634 \citep{ehlert11}, and the Phoenix cluster \citep{mcdonald12c}.
The median star formation rate for the 31 star-forming BCGs identified over the full redshift range is 50 M$_{\odot}$ yr$^{-1}$. It is worth noting that 5 of the 7 most star-forming BCGs (SFR $>$ 100 M$_{\odot}$ yr$^{-1}$) are at $z\ge0.9$, despite the fact that only 15\% of the clusters in this sample are at such high redshift. Likewise, 3 of the 5 most star-forming BCGs are at $z>1.1$, despite this representing 6\% of the sample.

We also show in the right panel of Figure \ref{fig:sfr_z} the average SFR as a function of redshift in five redshift bins. These averages, provided in Table \ref{table:ssfr}, have uncertainties that are derived by bootstrapping errors on individual data points. For non-detections, we assume zero SFR for the lower-bound of the uncertainty and the upper limit for the upper bound. Thus, the ranges shown encompass both the statistical uncertainty in the mean from detections and the uncertainty in the true value of the non-detections. This analysis reveals a strong evolution in the average star formation rate in BCGs over the past 9 Gyr. This increase in the average SFR is driven largely by a decrease in the number of non-detections at high-$z$, which we will discuss next.

\subsection{The Evolving Fraction of Star-Forming BCGs}

\begin{figure}[h!]
\centering
\includegraphics[width=0.48\textwidth,trim=0cm 0.25cm 0cm 0cm]{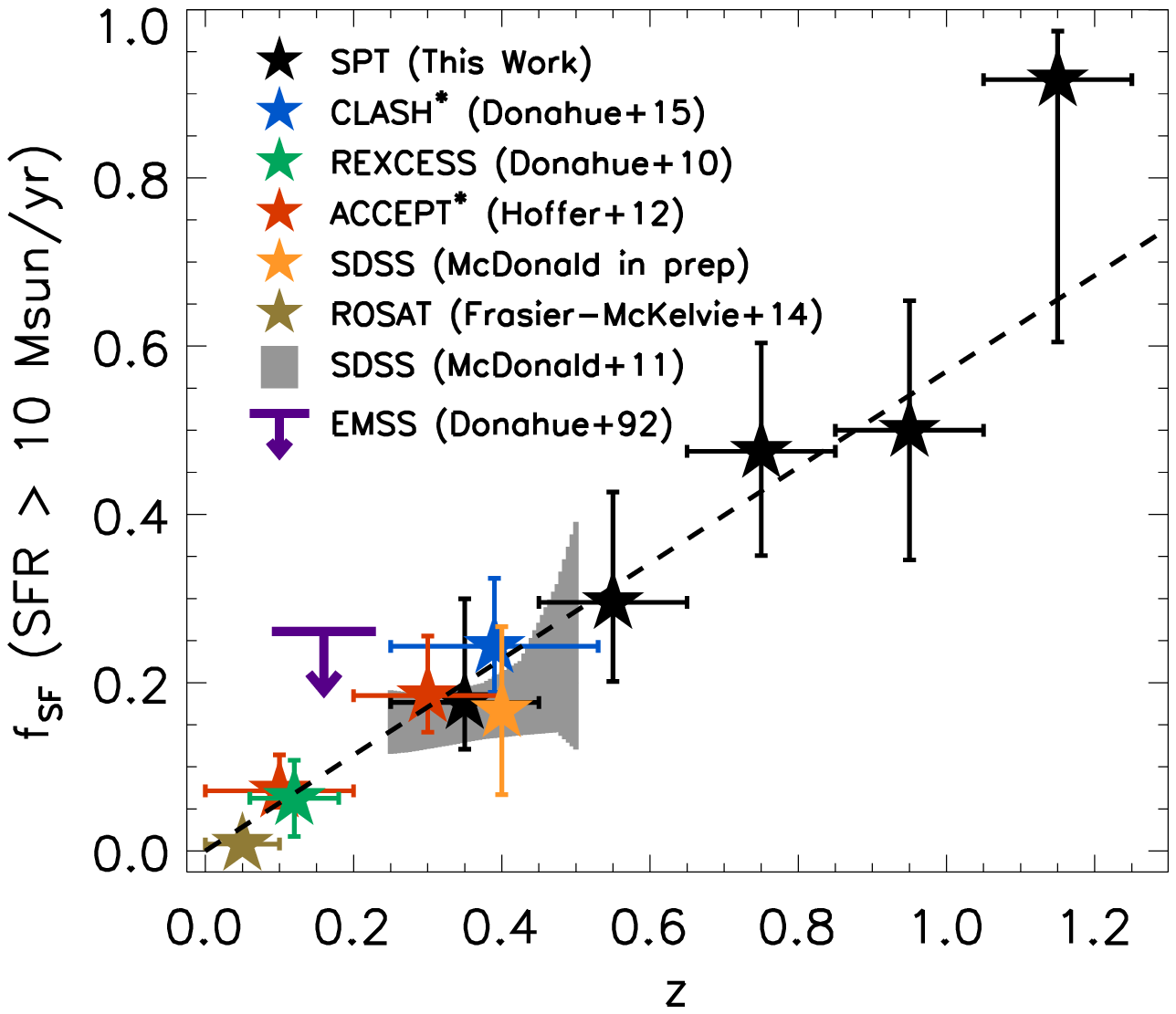}
\caption{Fraction of BCGs with SFR $> 10$ M$_{\odot}$ yr$^{-1}$ ($f_{SF}$) as a function of redshift. Black points show data from this work in four different redshift bins, while colored points show data from previous works over $0 < z < 0.5$ \citep{donahue92,donahue10,mcdonald11c, hoffer12, fraser14, donahue15}. Samples marked with an asterisk in the legend have had a bias correction applied, assuming that non-cool cores have passively-evolving BCGs and that the true underlying fraction of cool cores is 30\% (see \S3.2). This plot demonstrates consistency between a wide variety of samples based on X-ray, optical, and SZ selection. We find a steady rise in the fraction of star-forming BCGs, from $f_{SF}\sim0$\% at $z\sim0$ to $f_{SF}\sim70$\% at $z\sim1.2$. This growth is depicted with a best-fit dashed line. Vertical error bars are derived based on the methods described by \cite{cameron11} for binomial populations.}
\label{fig:sffrac}
\end{figure}

%\begin{figure}[htb]
%\centering
%\includegraphics[width=0.49\textwidth]{sffrac.eps}
%\caption{}
%\label{fig:sffrac}
%\end{figure}

In Figure \ref{fig:sffrac} we show how the fraction of BCGs with SFR $>$ 10 M$_{\odot}$ yr$^{-1}$ (hereafter $f_{SF}$) has evolved over time. Within the sample presented here, this fraction evolves from $\sim$20\% at $0.25 < z < 0.65$, to $\sim$50\% at $0.65 < z < 1.05$, to $\sim$90\% at $z>1.05$. This evolution is visible in Figure \ref{fig:sfr_z} as a relative lack of non-detections at the highest redshifts compared to the lowest redshifts. This result indicates that the fraction of clusters harboring a starburst in their central, most-massive galaxy grows by a factor of $\sim$3 between $z\sim0.6$ and $z\sim1.1$.

Even more intriguing is that, at $z\sim0$, $f_{SF} \sim 0-10$\% \citep{donahue10,fraser14}, suggesting an even more dramatic decline since $z>1$. We compile in Figure \ref{fig:sffrac} estimates of $f_{SF}$ from various surveys, spanning $0<z<0.5$. These surveys include both optically-selected \citep[][McDonald \etal in prep]{mcdonald11c} and X-ray-selected \citep{donahue92,donahue10,hoffer12,fraser14,donahue15} clusters, and are being compared here to an SZ-selected sample. For the ACCEPT \citep{cavagnolo09} and CLASH \citep{donahue15} samples, we apply a small correction due to the fact that both of these cluster samples are biased towards cool core clusters. This bias correction assumes that the true, underlying fraction of cool core clusters is 30\% \citep{haarsma10,hudson10, mcdonald13b} and that non-cool cores do not have star-forming BCGs at $z\lesssim0.5$ \citep{odea08,cavagnolo08,mcdonald10}. This well-motivated correction brings the observed value of $f_{SF}$ in line with other, more representative samples.
Encouragingly, at $z\sim0.4$ there are five distinct measurements of $f_{SF}$, based on three different selection methods, which all agree that the fraction of star-forming BCGs is $20\pm5$\%. The agreement between our measurement of $f_{SF}$ from SPT at $0.25 < z < 0.5$ and these earlier works suggests that our methodology is sound.

We show, in Figure \ref{fig:sffrac}, a linear fit to the star-forming BCG fraction, $f_{SF}$, as a function of redshift. The best-fitting line passes through the origin and has a slope of 0.57, implying a rapid evolution in $f_{SF}$ over the past $\sim$9 Gyr. Such a fit is not physically well-motivated, and does not capture the rapid growth at late times. However, regardless of the choice of parametrization, it is clear from Figure \ref{fig:sffrac} that the fraction of clusters harboring a strongly star-forming BCG at $z\sim1$ is \emph{significantly} ($>$3$\sigma$) higher than at $z\sim0$. This evolution is perhaps unsurprising, since galaxies in general were more star-forming at early times. To address this comparison, we next compare the BCG evolution to that observed in field galaxy and other cluster members over similar redshift intervals.
%In order to address why this may be, we next consider the properties of the host cluster for these star-forming systems.

\subsection{Specific Star Formation Rates of \\BCGs at $0.25 < z < 1.2$}

When comparing BCGs to other galaxies, it is necessary to normalize by the stellar mass of the individual galaxy. For a typical BCG, with M$_* \sim 10^{12}$ M$_{\odot}$, a SFR of 10 M$_{\odot}$ yr$^{-1}$ is negligible in terms of contributing to the overall mass of the cluster, requiring 100 Gyr to double the stellar mass. However, this same star formation rate in a low-mass galaxy like M82 is enough to power massive outflows and modify the galaxy morphology and stellar content on short (Myr) timescales \citep{forster03}. Thus, if we want to compare BCG evolution to field galaxy evolution, we must consider instead the \emph{specific} star formation rate, or sSFR, defined as sSFR $\equiv$ SFR/M$_*$.

\begin{deluxetable}{c c c c c}[htb]
\tablecaption{Average BCG Properties}
\tablehead{
\colhead{$z$} &
\colhead{N$_{cl}$} &
\colhead{$\left<M_*\right>$} &
\colhead{$\left<SFR\right>$} &
\colhead{$\left<sSFR\right>$}
\\
\colhead{} &
\colhead{} &
\colhead{[10$^{12}$ M$_{\odot}$] } &
\colhead{[M$_{\odot}$ yr$^{-1}$] } &
\colhead{[Gyr$^{-1}$]}
}
\startdata
%\begin{table}
%\centering
%\begin{tabular}{c c c c c}
%\hline\hline
%$z$ & N$_{cl}$ & $\left<M_*\right>$ & $\left<SFR\right>$ & $\left<sSFR\right>$\\
% & & [10$^{12}$ M$_{\odot}$] & [M$_{\odot}$ yr$^{-1}$] & [Gyr$^{-1}$]\\
% \hline
 0.25 $-$ 0.45& 19 & 2.09 & 4.0 $-$ 15 &0.002 $-$ 0.007\\
0.45 $-$ 0.65& 28 & 1.97 & 7.0 $-$ 24 &0.004 $-$ 0.012\\
0.65 $-$ 0.85& 22 & 1.47 & 20 $-$ 41 &0.013 $-$ 0.028\\
0.85 $-$ 1.05&  8 & 1.39 & 42 $-$ 84 &0.030 $-$ 0.061\\
1.05 $-$ 1.25&  6 & 1.54 & 89 $-$ 320 &0.058 $-$ 0.205
\enddata
\tablecomments{Average values for BCGs in five redshift bins, as shown in Figures \ref{fig:sfr_z} and \ref{fig:ssfr}. The ranges quoted on $\left<SFR\right>$ and $\left<sSFR\right>$ correspond to the combined uncertainty in the measurements (bootstrapping errors), the uncertainty in the BCG choice, and the added uncertainty associated with stacking non-detections. The total number of clusters is less than the number of BCGs due to the fact that we consider multiple potential BCGs for several clusters.}
\label{table:ssfr}
\end{deluxetable}

%\hline
%\end{tabular}
%\caption{Average values for BCGs in five redshift bins. The ranges quoted on $\left<SFR\right>$ and $\left<sSFR\right>$ correspond to the combined uncertainty in the measurements (bootstrapping errors), the uncertainty in the BCG choice, and the added uncertainty associated with stacking non-detections}
%\label{table:ssfr}
%\end{table}

\begin{figure}[htb]
\centering
\includegraphics[width=0.49\textwidth]{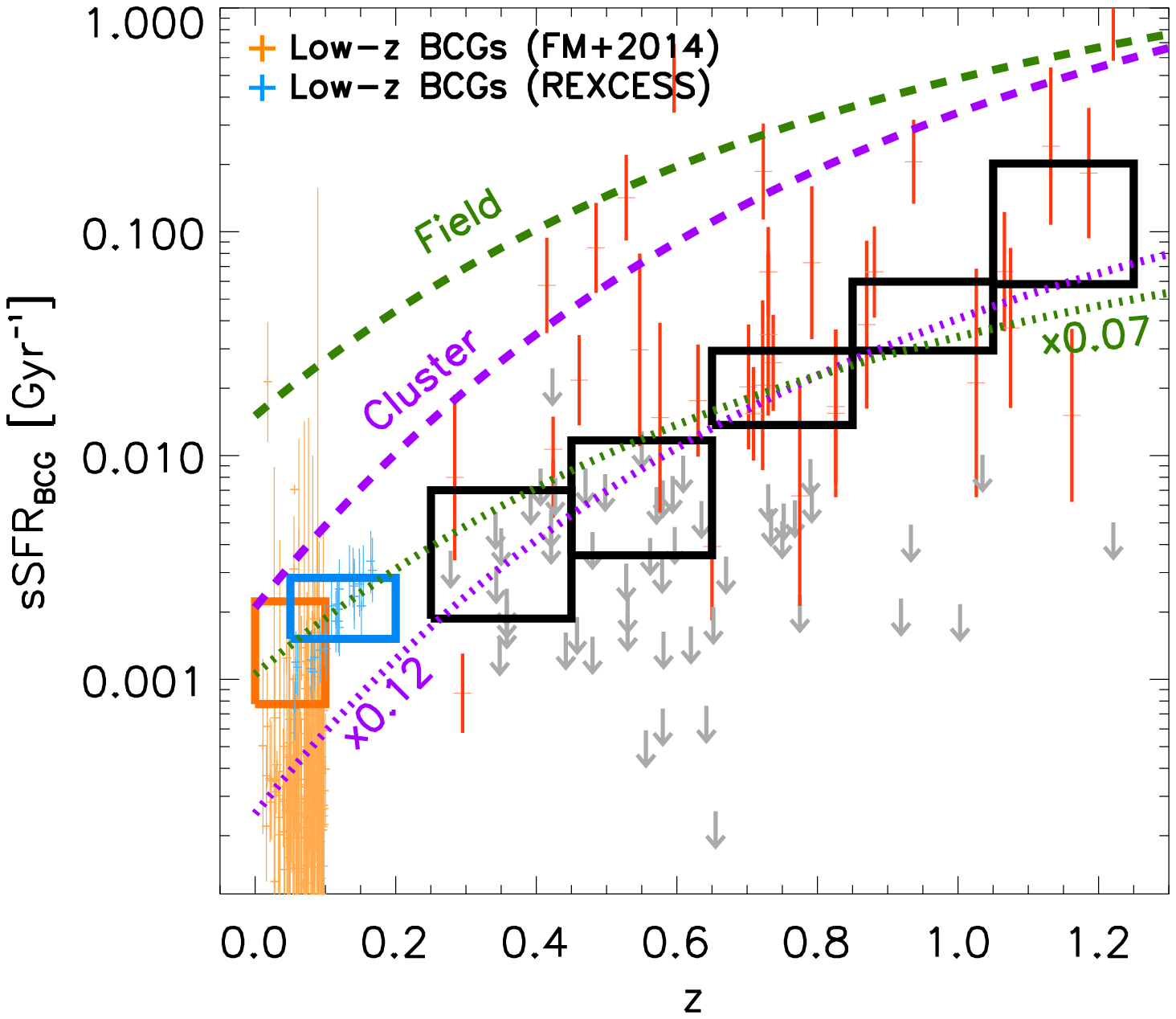}
\caption{Specific star formation rate (sSFR) as a function of redshift for BCGs in this work (red points, gray upper limits) and for nearby clusters \citep{haarsma10,fraser14}. Black boxes show the combined sSFR in different redshift bins, computed as the sum of the individual SFRs divided by the sum of stellar masses -- these points include the upper limits, with the vertical size of the boxes representing the associated uncertainty (see Table \ref{table:ssfr}). Dashed blue and purple curves show the evolution measured for the field and cluster environment, respectively, from \cite{alberts14}. We show these curves down, for comparison with the BCG evolution at $0.5 \lesssim z \lesssim 1$. At low-$z$, the BCG evolution is slower than for typical cluster members, suggesting an additional source of fuel that the non-BCG cluster galaxies are not accessing, while at high-$z$ ($z>1$) there appears to be a rapid increase in the amount of star formation in BCGs.
}
\label{fig:ssfr}
\end{figure}

In Figure \ref{fig:ssfr} we show the sSFR for BCGs in this work, as well as BCGs in similar-mass, low-$z$ clusters from \cite{haarsma10,fraser14}. In calculating the sSFR, we use the stellar mass of the old population obtained in the SED-fitting process (see \S2.3). For comparison, we show the average sSFR for field and cluster galaxies from \cite{alberts14}. This earlier work showed that, at present, cluster galaxies have suppressed star formation compared to the field. However, at $z\sim1.2$, galaxies in clusters are as star-forming as their field counterparts, and are evolving more rapidly. To match the analysis of \cite{alberts14}, who calculate sSFR(z) by stacking far-IR data on galaxy positions, we estimate average sSFR values for our data by separately summing the total SFR and total stellar mass for galaxies in five redshift bins. For BCGs with upper limits on their SFR, we assume two limiting cases: the case where the SFR is equal to the upper limit, and the case where the SFR is zero. The net result of this stacking is shown in Figure \ref{fig:ssfr} and Table \ref{table:ssfr}, where the height of the black boxes represents the combined uncertainty in the measurements (bootstrapping errors), the uncertainty in the BCG choice, and the added uncertainty associated with stacking non-detections.

Figure \ref{fig:ssfr} demonstrates that, at $0.5 \lesssim z \lesssim 1$, BCGs are evolving similarly to the cluster member galaxies suggesting a common fueling mechanism. At low-$z$ ($z\lesssim 0.5$), the evolution of BCGs is less rapid than in the cluster environment, suggesting that the quenching processes acting on member galaxies (e.g., ram pressure stripping, strangulation) may not be affecting the central galaxy. This change in slope may be due to the gas reservoir being replenished in BCGs by cooling of the ICM, which is likely fueling star formation in the lowest redshift BCGs \citep[e.g.,][]{mcdonald11b}. We will return to this idea in the discussion below. At $z\gtrsim1$ there is marginal evidence that BCGs may be evolving more rapidly than the member galaxies, suggesting a preferential quenching in the cluster core. 

\subsection{Which Clusters Host Star-Forming BCGs?}

\begin{figure}[htb]
\centering
\includegraphics[width=0.49\textwidth]{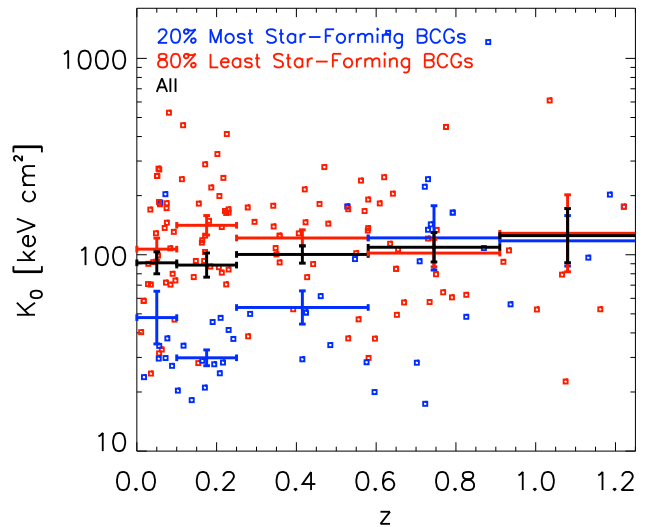}
\caption{Cluster central entropy (K$_0$) as a function of redshift for all clusters in this sample \citep[see also][]{mcdonald13b}. Here, we separate clusters into two subsamples within each redshift bin: those harboring the 20\% most star-forming BCGs (blue) and those harboring the 80\% least star-forming BCGs (red). At $0.0 < z < 0.25$ we show clusters from the ACCEPT database of \cite{cavagnolo09}, where star formation rates come from \cite{fraser14} ($z<0.1$) and \cite{crawford99} ($z>0.1$). For these low-$z$ clusters, we show only systems with M$_{500} \ge 3\times10^{14}$ M$_{\odot}$ and have recomputed the central entropy in an aperture with radius 15~kpc (corresponding to $\sim$0.01--0.02R$_{500}$), to prevent resolution bias \citep[see e.g.,][]{panagoulia13}.
In each redshift interval we show the mean K$_0$ and the error on the mean for both subsamples, as well as the total population (black). While at $z\lesssim0.6$ there is a clear separation in populations, with the most star-forming BCGs being found primarily in clusters with low-entropy cores \citep[e.g.,][]{cavagnolo08}, there appears to be no correspondence between cool cores at star-forming BCGs at $z\gtrsim0.6$.}
\label{fig:k0_sf}
\end{figure}

At low-$z$, star-forming BCGs are, with very few exceptions, found in low-entropy, relaxed ``cool core'' clusters \citep[e.g.,][]{cavagnolo08}. This appears to still be the case for CLASH clusters out to $z\sim0.4$ \citep{donahue15}, and for individual clusters at higher redshift \citep[e.g.][]{mcdonald12c}. However, it has not yet been established whether the presence of a cool core at high-$z$ correlates with a star-forming BCG.

\begin{figure}[htb]
\centering
\includegraphics[width=0.49\textwidth]{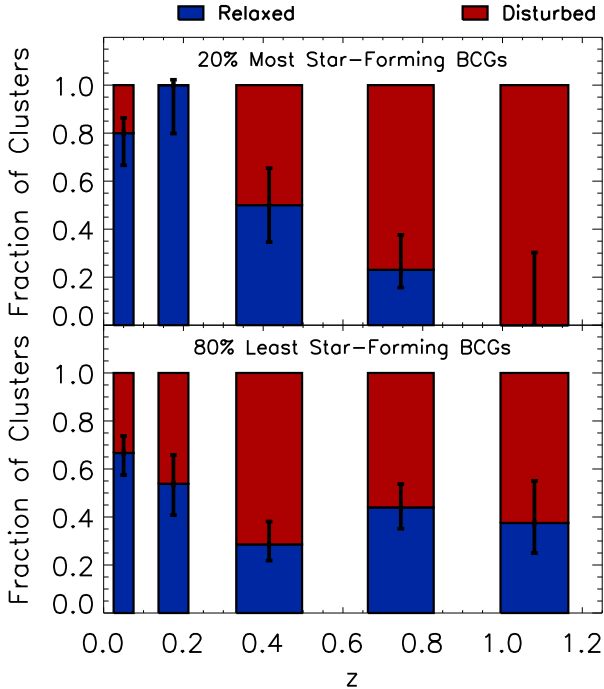}
\caption{This plot shows the fraction of relaxed (blue) and disturbed (red) clusters as a function of redshift, for two subsamples: clusters harboring the 20\% most star-forming BCGs (upper panel) and those harboring the 80\% least star-forming BCGs (lower panel). For low-redshift clusters ($z < 0.3$, two leftmost bins) relaxedness has been determined based on the X-ray ``symmetry'' reported in \cite{mantz15}, while star formation rates come from \cite{fraser14} and \cite{crawford99}. For the three higher-redshift bins, relaxedness is quantified following \cite{nurgaliev13}. This figure demonstrates that the most star-forming BCGs tend to be found in relaxed clusters at low-$z$, in agreement with the literature, while at high-$z$ they are found predominantly in morphologically disturbed clusters. There is no strong trend in the the morphology of clusters hosting the 80\% least star-forming BCGs.}
%Fraction of BCGs that are star-forming (blue, SFR~$>$~10~M$_{\odot}$~yr$^{-1}$) or passive (red, SFR~$<$~10~M$_{\odot}$~yr$^{-1}$) in relaxed and unrelaxed clusters, as a function of redshift. For low-redshift clusters ($z<0.3$, two leftmost bins), relaxedness has been determined based on the X-ray ``symmetry'' reported in \cite{mantz15}, while star formation rates come from \cite{fraser14} and \cite{crawford99}. For the three higher-redshift bins, relaxedness is quantified following \cite{nurgaliev13}. Relaxed clusters, shown in the top panel, tend to harbor star-forming BCGs in 25--50\% of cases, independent of redshift. On the contrary, unrelaxed clusters have an evolving star-forming fraction, from $\sim$0\% at $z\sim0$, to $\sim$80\% at $z\sim1$. This suggests that the mechanism responsible for star formation in high-$z$ BCGs is linked to the dynamical state of the cluster.}
\label{fig:morph}
\end{figure}

In Figure \ref{fig:k0_sf}, we show the distribution of central entropy (K$_0$) as a function of redshift for clusters hosting the most star-forming BCGs compared to those hosting the most passive BCGs. Specifically, in each redshift bin we separately compute the average central entropy for the clusters harboring the 20\% most star-forming BCGs (e.g., Perseus, Abell\,1835, Phoenix, etc) and for those harboring the 80\% most passive BCGs. At $z<0.3$, we show clusters from the ACCEPT sample \citep{cavagnolo09}, where the SFRs are inferred from archival infrared data \citep[$z<0.1$;][]{fraser14} or the H$\alpha$ line luminosity \citep[$z>0.1$;][]{crawford99}. We have recomputed the central entropy in a 15~kpc aperture ($\sim$0.015R$_{500}$) for clusters at $z<0.3$ to match our more coarse, high-$z$ aperture, and to avoid resolution bias \citep[e.g.,][]{panagoulia13}. We include only clusters with M$_{500} > 3\times10^{14}$ M$_{\odot}$ \citep[based on L$_X$--M relation from][]{pratt09}, to mimic the SPT selection. We find, in agreement with previous works \citep{cavagnolo08},  that clusters hosting the most star-forming BCGs have a typical central entropy of K$_0 \sim 30$ keV cm$^2$, while those hosting the least star-forming BCGs have K$_0 \sim 100$ keV cm$^2$. As the redshift increases, the average core entropy in clusters hosting the most star-forming BCGs increases to $\sim$40 keV cm$^2$ ($z\sim0.4$), and then to $\sim$100 keV cm$^2$ ($z\sim0.7$).  In the two highest-redshift bins ($z\gtrsim0.6$), there is no statistical difference between the distribution of core entropy in clusters with star-forming and passive BCGs. We find no strong evolution in the core entropy of the full sample, or in the subsample of clusters with passively-evolving BCGs.
This seems to suggest that, while ICM cooling is likely responsible for providing the fuel for star formation in low-$z$ clusters, a different mechanism is responsible for star formation in the high-$z$ BCGs.

% What about morphology?

Another mechanism for forming stars in BCGs is via mergers with gas-rich galaxies. Under the assumption that such mergers happen most often shortly after the infall of a group or other massive halo, which provides an influx of new galaxies along orbits that may not be stable, we would expect BCGs with star formation being fueled by mergers to reside in clusters with disturbed X-ray morphology. 
In Figure \ref{fig:morph}, we test this scenario. For clusters in the SPT sample, we define X-ray morphology using the ``$a_{phot}$'' parameter, following \cite{nurgaliev13}, with a relaxed critereon of $a_{phot} < 0.1$. This quantity is less biased to signal-to-noise than other indicators, such as power ratios and centroid shift, but is consistent with these in the limit of high signal to noise, as is demonstrated by \cite{nurgaliev13}. At low-$z$, we use the recently-compiled list of ``symmetry'' measurements from \cite{mantz15}, using overlapping clusters to determine a common ``relaxed'' criterion. As before, for clusters at $z<0.1$ and $0.1 < z < 0.3$, we use BCG star formation rates from \cite{fraser14} and \cite{crawford99}, respectively, cutting on mass (M$_{500} \ge 3\times10^{14}$ M$_{\odot}$) in order to ensure uniformity in the samples.

Figure \ref{fig:morph} confirms that, at $z<0.3$, the most star-forming BCGs tend to reside in relaxed, cool core clusters. The star formation in these BCGs is most likely being fueled by the cooling ICM, where cooling flows are most commonly found in relaxed clusters. On the contrary, at $z\gtrsim$0.6, the most star-forming BCGs are found more often in clusters with disturbed X-ray morphology, with $\sim$90\% of the most star-forming BCGs at $z\gtrsim0.6$ being found in such systems. This implies that, at early times, star formation in the BCG is more strongly correlated with the dynamic state, rather than the cooling state, of the ICM -- the inverse to what is observed in the nearby Universe.

Figures \ref{fig:k0_sf} and \ref{fig:morph} suggest that there is a transition between star formation in low-$z$ BCGs being linked to low-entropy, cool core clusters, to star formation in high-$z$ BCGs being related more to a disturbed cluster morphology than to the cooling properties of the ICM. Below we will discuss possible interpretations of this result, and others presented thus far, while also addressing any potential systematic biases in this study.

\section{Discussion}

\begin{figure*}[h!]
\centering
\includegraphics[width=0.96\textwidth,trim=0.cm 0.3cm 0cm 0.3cm,clip=true]{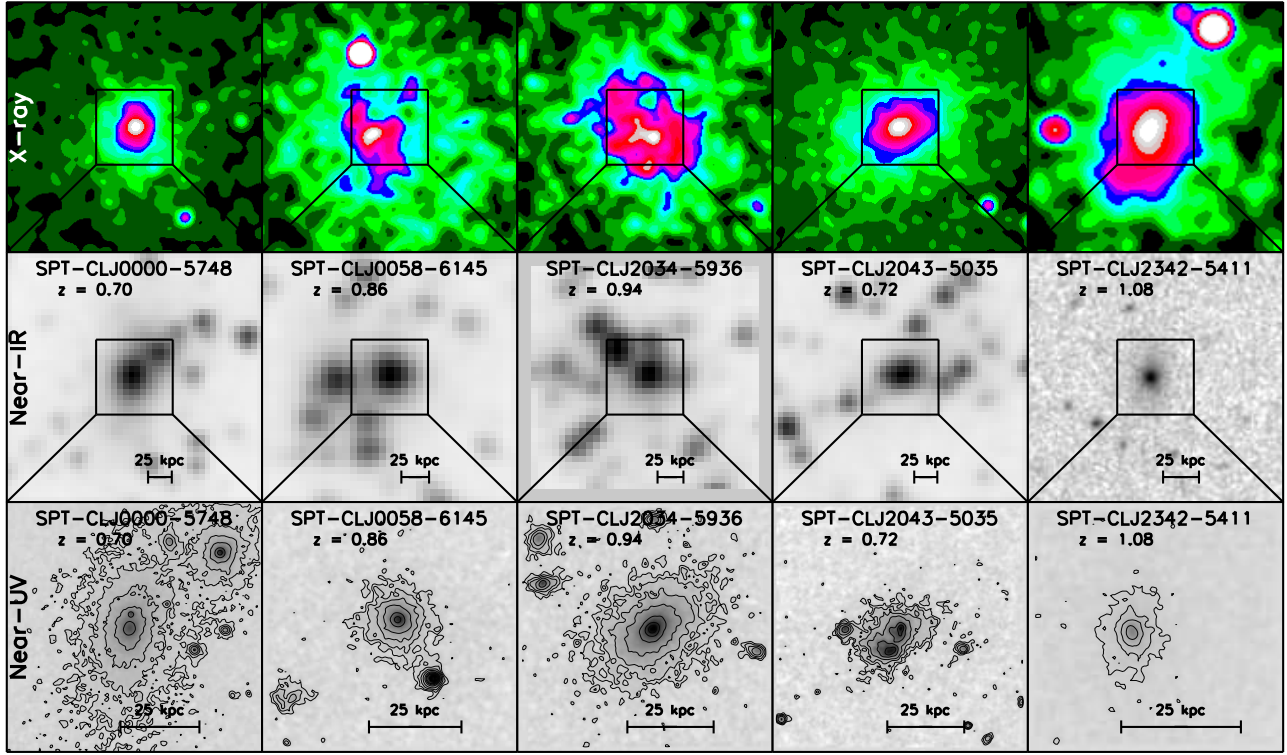}
\caption{\emph{Upper row}: Smoothed X-ray images from \emph{Chandra} of five clusters from this paper at $z>0.7$ with rest-frame near-UV imaging from HST and with both relaxed X-ray morphology and a low-entropy core. The field of view for each cluster corresponds to R$_{500}$ on a side. \emph{Middle row}: Near-infrared image (FourStar or \emph{Spitzer}) of the inner region of the cluster. The footprint of this image is overlaid on the X-ray image above. In all panels, the galaxy identified as the BCG is at the center of the field of view. \emph{Lower row}: Rest-frame near-UV image of the BCG from HST. The footprint of each image is shown on the near-IR image directly above it. These images show that, for the most part, the relatively smooth UV emission in these galaxies is tracing the underlying old stellar populations, contrary to the highly-asymmetric, filamentary star formation observed in the cores of low-$z$, relaxed galaxy clusters..}
\label{fig:highz_cc_thumbs}
\end{figure*}

\begin{figure*}[h!]
\centering
\includegraphics[width=0.96\textwidth,trim=0.cm 0.3cm 0cm 0.3cm,clip=true]{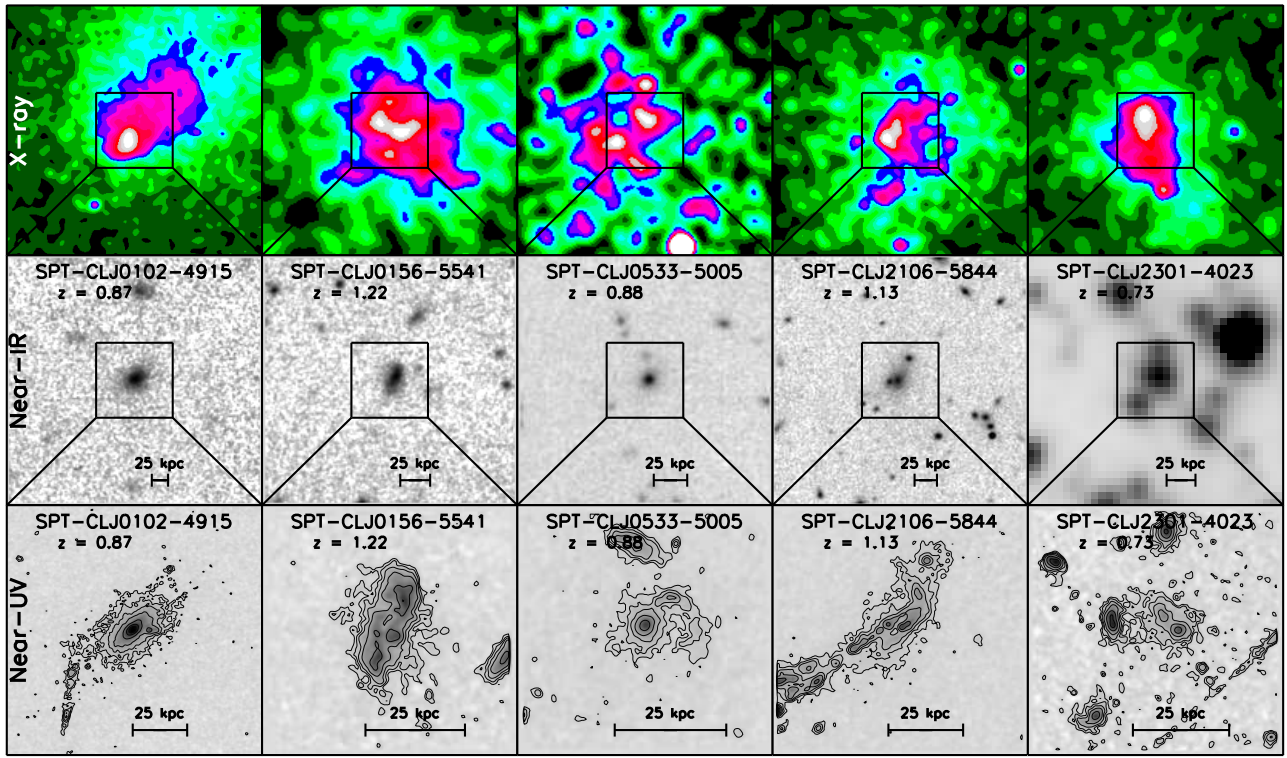}
\caption{Similar to Figure \ref{fig:highz_cc_thumbs}, but now showing five clusters at $z>0.7$ with disturbed X-ray morphologies. The BCGs in these clusters all have highly asymmetric UV emission, suggesting a different mode of star formation than the BCGs in relaxed clusters.}
\label{fig:highz_ncc_thumbs}
\end{figure*}

\subsection{Comparing X-ray and UV Morphology for Individual Star-Forming BCGs}

Based on Figures \ref{fig:k0_sf} and \ref{fig:morph}, there appears to be an evolving connection between star formation in the BCG and the host cluster morphology. In order to investigate the fuel source for this star formation (i.e., cooling, gas-rich mergers, etc), and determine whether this is linked to the dynamical state of the cluster, we require rest-frame UV imaging at significantly higher angular resolution.
Fortunately, many of the high-$z$ clusters in this sample have been observed by the \emph{Hubble Space Telescope} (HST) as part of various weak-lensing programs (PIs: Stubbs, High, Schrabback). These observations are all in the F606W filter, which corresponds to rest-frame $u$-band at $z\sim0.7$ and $<$3000\AA\ at $z>1$. These data provide a detailed view of the young stellar populations in a subsample (10/14) of high-$z$ star-forming BCGs. 

In Figures \ref{fig:highz_cc_thumbs} and \ref{fig:highz_ncc_thumbs} we highlight the X-ray and UV morphology of the cluster and BCG, respectively, for all 10 clusters with available rest-frame $u$-band (or bluer) HST imaging at $z>0.7$. For each cluster, we also show a near-IR image of the core region, to demonstrate that these star-forming galaxies do indeed have the highest stellar mass of all the galaxies in the core, consistent with being the BCG.
We have divided these 10 clusters by X-ray morphology into two subsamples of relaxed ($a_{phot} < 0.3$ and K$_0 < 95$ keV cm$^2$) and unrelaxed ($a_{phot} > 0.3$ or K$_0 > 95$ keV cm$^2$). 
The five most relaxed clusters (Figure \ref{fig:highz_cc_thumbs}), in general, have relatively smooth, symmetric UV morphologies that resemble the underlying old stellar (near-IR) distribution. We caution that much of this near-UV emission may originate in the old stellar populations \citep[see e.g.][]{hicks10} -- we require either far-UV imaging or equally high angular resolution imaging in a near-IR band (allowing subtraction of the old population) to determine the morphology of the excess UV emission due exclusively to young stars. Two of the BCGs shown in Figure \ref{fig:highz_cc_thumbs} (SPT-CLJ0000-5748, SPT-CLJ2043-5035) appear to be in the midst of major mergers, based on both the near-IR and near-UV imaging, despite the fact that these two BCGs reside in the two most relaxed high-$z$ clusters in our sample. None of the star-forming BCGs in this high-$z$, relaxed cluster subsample show evidence for extended, asymmetric filaments of star formation which are commonly found in analogous low-$z$ systems \citep[e.g.,][Tremblay \etal 2015]{odea10,mcdonald11b,donahue15}. Instead, the UV emission is concentrated in the BCG center for 4 out of 5 BCGs, perhaps indicating that these systems are experiencing either nuclear starbursts or are AGN misidentified as star-forming galaxies.

On the other hand, in the five least relaxed clusters (Figure \ref{fig:highz_ncc_thumbs}) the UV emission in and around the BCG is clumpy and filamentary. In particular, SPT-CLJ0102-4915 and SPT-CLJ2106-5844 exhibit clumpy UV emission extending 61~kpc and 57~kpc from the BCG center, respectively. This is comparable to the most extended star-forming filaments found in Abell~1795 \citep[50~kpc;][]{mcdonald09} and Perseus \citep[60~kpc;][]{conselice01,canning14}. In all five of these BCGs, there are a minimum of two distinct UV peaks -- in several systems, the brightest UV peak is offset from the BCG nucleus. Unlike the five BCGs in the more relaxed clusters, none of these systems are consistent with being purely AGN, given their complex morphology.

While incomplete, these HST data suggest a qualitative difference in UV morphology between BCGs in relaxed, cool-core clusters and those in unrelaxed systems. In relaxed, cool-core clusters, where low-$z$ surveys tend to find star-forming BCGs, we find (in 4/5 cases) smooth, centrally concentrated UV morphologies. In the more morphologically-disturbed systems (based on X-ray imaging), we find (in 5/5 cases) clumpy or filamentary UV emission, with several ($\gtrsim$3) distinct emission peaks. Follow-up studies involving deep, high spatial resolution imaging at both far-UV and near-IR of a larger sample of high-$z$ BCGs are necessary to determine whether this emerging trend is merely coincidence or evidence for a link between central star formation and the dynamical state of the cluster core.

\subsection{An Evolving Fuel Supply?}

The results presented in \S3 suggest that, at high-$z$, the potential for star formation in the BCG is maximized when that BCG belongs to a dynamically-active cluster. This is contrary to the established wisdom, based on numerous studies of low-$z$ clusters, that the most relaxed, cool core clusters tend to harbor the most star-forming BCGs \citep[e.g.,][]{crawford99,edwards07,cavagnolo08, donahue10}.  We proceed with a discussion of these results under the assumption that the X-ray morphology traces the dynamical state of the cluster, or, more specifically, that morphologically-disturbed clusters in the X-ray are undergoing (or have recently undergone) a major merger. 

%\subsubsection{An Increase in the Rate of Gas-Rich Mergers}

At early times, the cores of galaxy clusters contained a higher fraction of star-forming galaxies than they do today -- this is known as the ``Butcher-Oemler Effect'' \citep{butcher84}. This trend continues with increasing redshift, such that at $z\sim1.4$, the star formation rate in the field and in the cluster environment are indistinguishable \citep[][see also Figure \ref{fig:ssfr}]{brodwin13,alberts14,wagner15}. Thus, we would expect that there is a significantly higher fraction of gas-rich galaxies in our high-$z$ subsample than in the low-$z$ systems, particularly in the inner cores where gas depletion is most effective at low-$z$. Regardless of the impact angle of a cluster-cluster merger (as long as it is a bound system), the cores will typically pass through one another within $\sim$2 crossings \citep[e.g.,][]{ricker01,poole06}. The crossing of these two cores will result in both an increase in the effectiveness of ram-pressure stripping, due to the high relative velocities of the cores, and an increased rate of galaxy-galaxy mergers and harassment. Each of these elevated processes result in the removal of cool gas from member galaxy halos. Assuming that this gas remains cool, this would lead to an increase in the availability of fuel for star formation in the cores of merging clusters. This is not the case in low-$z$ systems, since galaxies in the cores of low-$z$ clusters tend to be gas-poor (hence the low star formation rates), so very little gas is available to be removed from their halos and contributed to the core.

\begin{figure}[htb]
\centering
%\begin{tabular}{c}
%\includegraphics[width=0.48\textwidth]{sffrac_fit1.eps}
%\includegraphics[width=0.48\textwidth,trim=1cm 0cm 0cm 0cm]{sffrac_fit2.eps}\\
\includegraphics[width=0.48\textwidth,trim=1cm 0cm 0cm 0cm]{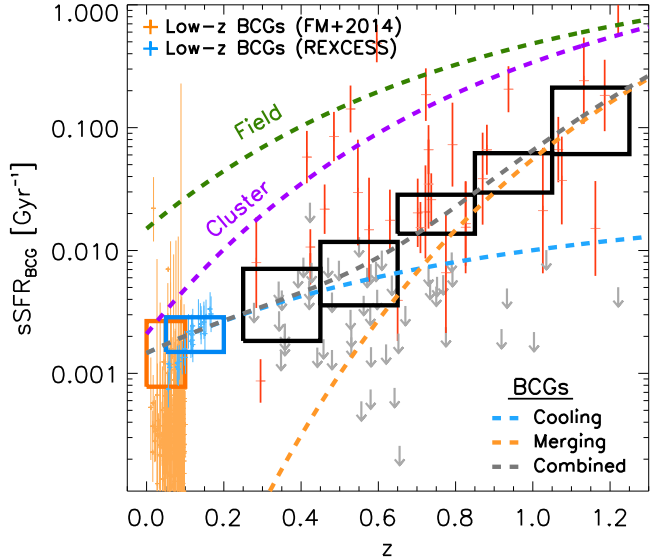}
%\end{tabular}
\caption{Here, we repeat Figure \ref{fig:ssfr}, now with a two-component fit to the BCG evolution. This fit assumes two epochs of star formation in BCGs: low-level star formation at present day, which correlates with the presence of a relaxed morphology and cool core in the ICM, and a rapid increase in star formation at $z\gtrsim0.8$, which correlates with the presence of disturbed ICM morphology. We propose that this is indicative of two epochs of BCG star formation: merger-driven at early times, and cooling-induced at late times.}
\label{fig:sffr_v2}
\end{figure}

With this scenario in mind, we reexamine Figure \ref{fig:ssfr}, modeling the observed evolution in BCG star formation with two distinct evolutionary components. These models, shown in Figure \ref{fig:sffr_v2}, assume that the evolution can be described as the sum of two power laws in time ($\left<sSFR\right> \propto e^{t/\tau}$). The best-fitting models have decay times of 4 Gyr (low-$z$) and 0.7 Gyr (high-$z$), compared to 1.5 Gyr and 2.2 Gyr for high-$z$ cluster and field galaxies, respectively, from \cite{alberts14}. This model is meant to provide a qualitative assessment of what may be driving star formation in BCGs over time -- there is no shortage of models with an equal number of free parameters that would provide an equivalently good fit to these data. 

The models shown in Figure \ref{fig:ssfr} were chosen to illustrate two epochs of declining star formation. From $z\sim1.4$ to $z\sim0.8$, the sSFR declines more rapidly ($\tau = 0.7$ Gyr) than both the field ($\tau = 2.2$ Gyr) and cluster ($\tau = 1.5$ Gyr) environment, implying that the BCG is quenched more rapidly than the other cluster members. This is consistent with an extrapolation from the field to the cluster environment from \cite{alberts14}, indicating that (unsurprisingly) quenching of star formation is a strong function of local galaxy density. At late time, from $z\sim0.6$ to $z\sim0$, the sSFR of BCGs is evolving more slowly ($\tau = 4$ Gyr) than both the field and cluster environment. Considering that field galaxy evolution is approximately passive, this implies that BCGs have an additional source of fuel -- presumably the cooling ICM in relaxed systems. The non-zero slope of this evolution may be due to an ever-improving balance between ICM cooling and AGN feedback in the cores of galaxy clusters, leading to a reduction in the efficiency of ICM cooling over time.

In summary, there appears to be a transition in the source of fuel for star formation in BCGs over the past $\sim$10 Gyr. In nearby clusters, star formation is likely to be fueled by the cooling ICM, and regulated by feedback from the central AGN. At early times, the most star-forming BCGs are found in dynamically unrelaxed clusters, suggesting that star formation may be predominantly fueled by interactions with other galaxies, similar to the other, non-BCG members.

\subsection{Lowering the Precipitation Threshold in Mergers}
Several recent studies have suggested that the condition for thermal instability in the hot ICM can be described as $t_{cool}/t_{ff} = 10$, where $t_{cool}$ is the cooling time, and $t_{ff}$ is the local free-fall time \citep{mccourt12, sharma12, gaspari12, voit15a,voit15b}. This threshold can be interpreted as the ratio of the cooling time to the mixing time. Assuming that AGN feedback is anisotropic, preventing local thermal instabilities requires that the heated gas mixes with the cooling gas on timescales shorter than the local cooling time.

In order to cross this threshold, one of two conditions can be met. The gas can cool, reducing the local cooling time, $t_{cool}$. In the precipitation-driven feedback scenario of \cite{voit15b}, this leads to rapid precipitation of cold clouds, which fuel AGN feedback, leading to an increase of $t_{cool}$. This cycle can repeat indefinitely, until a more energetic process drives $t_{cool}$ to much larger values ($\gg$1 Gyr). Alternatively, if the free-fall time is increased, gas at a fixed density and temperature will more readily condense out of the hot phase. This criteria is, in principle, met if the dense core of a galaxy cluster is dislodged from the minimum of the dark matter potential, as is the case during an interaction with another massive group or cluster. It remains unclear, however, whether increasing the free-fall time of the gas in this way would, in fact, lead to more favorable cooling. Strictly speaking, a dislodged cool core would have a lower value of $t_{cool}/t_{ff}$, but the simulations which arrived at the threshold of $t_{cool}/t_{ff} = 10$ for thermal instability assumed relaxed clusters. In the case of a relaxed cluster, $t_{ff}$ is a proxy for the convective timescale of the hot gas. However, in a merging cluster, bulk flows will likely drive mixing on faster timescales, perhaps leading to \emph{less} efficient cooling.

%\vskip 0.1in
This represents an alternate scenario which would lead to a link between enhanced star formation in the BCG and an unrelaxed dynamical state of the cluster core at early times. We lean toward the merger-induced SF scenario to explain the uptick in star-forming BCGs at early times, but stress that follow-up studies, including spatially-resolved far-UV (e.g., HST) and far-IR (e.g., ALMA) imaging and optical spectroscopy, of these BCGs are necessary to provide additional insights into the mechanism for enhancing star formation in these high-$z$ BCGs.

\subsection{AGN Contamination}
As we discussed in \S2.6,  it is possible that many of the BCGs that we have identified as star-forming may, instead, be AGN. With the inhomogeneous data set in hand, it is challenging to differentiate between starburst and AGN, or composite systems, in a uniform way for the full sample. Thus, it may be that both the estimated fraction of star-forming BCGs and the absolute star formation rates of these BCGs may be biased high. 

Considering only clusters at $z>0.75$ (the two highest redshift bins in Figure \ref{fig:sffrac}) we detect star formation in 9--10 out of 22 (41--45\%) of BCGs, where the uncertainty represents cases where multiple BCGs were identified in a given cluster. This is a factor of two higher than the fraction of star-forming BCGs at $z\sim0.4$. If we consider only the subsample of clusters with HST follow-up, and label systems with symmetric, nuclear UV emission as non-star forming (see Figures \ref{fig:highz_cc_thumbs} and \ref{fig:highz_ncc_thumbs}), we find evidence of extended, asymmetric star formation in 4 of 10 (40\%) of BCGs. Given the reduction in sample size by requiring HST follow-up, the uncertainty on this fraction is larger, with the 95\% confidence interval being 16.7--69.2\%. While this is consistent with the observed star-forming fraction of $\sim$18\% at $z\sim0.4$ \citep{mcdonald11c, donahue15}, it is inconsistent with the lower value of $\sim$5\% measured at $z\sim0$ \citep{donahue10,fraser14}. Thus, while we can say with high confidence that the fraction of star-forming BCGs has evolved significantly from $z\sim0$ to $z\gtrsim0.4$, our inability to differentiate between central AGN and nuclear starbursts limits our ability to say with certainty whether this fraction continues to grow.

It is important to note, however, that the fraction of nearby BCGs harboring AGN that are bright in the UV or IR continuum is low -- the vast majority of cluster-centric AGN are radio galaxies which appear quiescent at most other wavelengths. If the high fraction of systems identified here as star-forming are instead AGN, it implies a high accretion rate onto the supermassive black hole at early times \citep[e.g.,][]{russell13}.
Of the hundreds of known BCGs at low-$z$, there are only a handful of systems that appear to harbor rapidly-accreting, radiatively-efficient AGN \citep[e.g.,][]{russell10,osullivan12,ueda13,kirk14,reynolds14,walker14}. Thus, we can say with confidence that there is a dramatic increase in the amount of ``activity'' in BCGs from $z\sim0$ to $z\sim1$ -- whether that activity refers to massive bursts of star formation or the rapid growth of central supermassive black holes remains an open question.

\section{Summary}
We present multiwavelength observations and inferred star formation rates for 90 brightest cluster galaxies (BCGs) in SPT-selected galaxy clusters at $0.25 < z < 1.25$, all of which have archival X-ray data. The main results from this study can be summarized as follows:

\vskip 0.1in
\noindent{}$\bullet$~~We find a significant number of BCGs (31/90) with SFR $>$ 10 M$_{\odot}$ yr$^{-1}$, representing a much higher occurrence rate than that observed in galaxy clusters at $z\sim0$ \citep[$\sim$1--5\%;][]{donahue10,fraser14}. Of these 31 BCGs, one quarter (7/31) have SFR $>$ 100 M$_{\odot}$ yr$^{-1}$.

\vskip 0.1in
\noindent{}$\bullet$~~The fraction of clusters harboring a star-forming (SFR $>$ 10 M$_{\odot}$ yr$^{-1}$) BCG is found to be $\sim$20\% at $z\sim0.4$, consistent with many earlier works based on optical and X-ray cluster selection. This fraction rises rapidly at $z\gtrsim0.8$, to a measured value of $\sim$90\% at $z\sim1.1$.

\vskip 0.1in
\noindent{}$\bullet$~~The specific star formation rate ($sSFR \equiv SFR/M_*$) of BCGs has evolved more slowly ($\tau = 4$ Gyr) in recent times ($z\lesssim0.6$) than the overall cluster ($\tau = 1.5$ Gyr) and field ($\tau = 2.2$ Gyr) populations. This is most likely due to a replenishment of gas in the BCG via cooling of the ICM. At early times, the evolution was more rapid ($\tau = 0.7$ Gyr), with the sSFR in BCGs dropping from $\sim$0.1 Gyr$^{-1}$ at $z =1.2$ to $\sim$0.02 Gyr$^{-1}$ at $z=0.8$.

%\vskip 0.1in
%\noindent{}$\bullet$~~The rise in the fraction of star-forming BCGs with increasing redshift appears to be primarily driven by an increase in the fraction of star-forming BCGs in unrelaxed, non-cool core clusters. Such systems tend to harbor passively evolving BCGs at $z\sim0$, while at $z\sim1$ they harbor some of the most star-forming BCGs known.

\vskip 0.1in
\noindent{}$\bullet$~~At $z\gtrsim0.6$ there is no significant correlation between the central entropy of the host cluster and the presence of star formation signatures in the BCG, contrary to what is observed in nearby clusters.

\vskip 0.1in
\noindent{}$\bullet$~~While, at $z\sim0$, star-forming BCGs are found in the centers of relaxed, cool core clusters, this trend appears to reverse at high-$z$. At $z\gtrsim0.6$, the most star-forming BCGs in this sample are found in the cores of morphologically-disturbed clusters (based on X-ray asymmetry).

\vskip 0.1in
\noindent{}$\bullet$~~Excluding the Phoenix cluster, the most strongly star-forming systems in this sample have SFRs of order $\sim$100--300 M$_{\odot}$ yr$^{-1}$. Based on rest-frame near-UV follow-up of a subsample of high-$z$ systems with HST, we find that this star formation can be extended on scales of $\sim$50--60 kpc.

\vskip 0.1in
The observation that an enhancement in BCG star formation correlates with the dynamical state of the cluster at high-$z$ suggests that star formation may have been fueled by interactions with gas-rich satellites at early times. Further studies, utilizing deep, high angular resolution far-IR and far-UV imaging and integral field spectroscopy of a larger sample of BCGs will help determine if such mergers are, indeed, the dominant source of star formation in BCGs at early times. 

\section*{Acknowledgements} 
We thank Mark Voit and John ZuHone for helpful conversations.
M. M. acknowledges support by NASA through contracts HST-GO-13456.002A (Hubble) and GO4-15122A (Chandra), and Hubble Fellowship grant HST-HF51308.01-A awarded by the Space Telescope Science Institute, which is operated by the Association of Universities for Research in Astronomy, Inc., for NASA, under contract NAS 5-26555. 
The South Pole Telescope program is supported by the National Science Foundation through grants ANT-0638937 and PLR-1248097. 
Partial support is also provided by the NSF Physics Frontier Center grant PHY-0114422 to the Kavli Institute of Cosmological Physics at the University of Chicago, the Kavli Foundation, and the Gordon and Betty Moore Foundation. 
Support for X-ray analysis was provided by NASA through Chandra Award Numbers 12800071, 12800088, and 13800883 issued by the Chandra X-ray Observatory Center, which is operated by the Smithsonian Astrophysical Observatory for and on behalf of NASA. 
Galaxy cluster research at Harvard is supported by NSF grant AST-1009012 and at SAO by NSF grants AST-1009649 and MRI-0723073. 
The McGill group acknowledges funding from the National Sciences and Engineering Research Council of Canada, Canada Research Chairs program, and the Canadian Institute for Advanced Research.
Argonne National Laboratory's work was supported under U.S. Department of Energy contract DE-AC02-06CH11357. 
This work is based in part on observations made with the Spitzer Space Telescope, which is operated by the Jet Propulsion Laboratory, California Institute of Technology under a contract with NASA.
J.E.C. acknowledges support from National Science Foundation grants PLR-1248097 and PHY-1125897.
CR acknowledges support from the University of Melbourne and from the Australian Research CouncilÕs Discovery Projects scheme (DP150103208).
DA and TS acknowledge support from the German Federal Ministry of Economics and Technology (BMWi) provided through DLR under projects 50 OR 1210,  50 OR 1308,  and 50 OR 1407. 

\vskip 0.1in
\noindent{}{\it Facilities:} 
\facility{Blanco (MOSAIC)},
\facility{CXO (ACIS)},
\facility{GALEX},
\facility{HST (ACS)},
\facility{Magellan:Baade (FourStar, IMACS)}, 
\facility{Magellan:Clay (LDSS3, Megacam)}, 
\facility{Max Planck:2.2m (WFI)},
\facility{NTT (EFOSC)},
\facility{Spitzer (IRAC)},
\facility{Swope (SITe3)}, 
\facility{VLT:Antu (FORS2)},
\facility{WISE},
\facility{XMM (OM)}

%\bibliographystyle{apj}
%\bibliography{ref}

% ===== APPENDIX ====== %

%\clearpage

\renewcommand\thefigure{\thesection.\arabic{figure}}    
\setcounter{figure}{0}    
\renewcommand\thetable{\thesection.\arabic{table}}    
\setcounter{table}{0}    

\begin{appendices}
\section{Spectral Energy Distributions}
%
%\begin{figure*}[h!]
%\centering
%\includegraphics[width=0.99\textwidth]{sedplot.eps}
%\caption{Observed broadband spectral energy distribution for 90 BCGs in this sample. We overplot the best-fitting model, which consists of an old stellar population which formed at $z=2$ \citep[red;][]{leitherer99}, a young stellar population which has been continuously forming stars for 100 Myr \citep[blue;][]{leitherer99}, and a dust component parametrized by a powerlaw in the mid-IR \citep[purple;][]{casey12}. When a young or dusty component was not detected, we show the upper limit on the UV and/or IR continuum with a dotted line.}
%\label{fig:seds}
%\end{figure*}
%
\begin{figure*}[h!]
\centering
\includegraphics[width=0.99\textwidth,trim=0cm 3.5cm 0cm 0cm,clip=true]{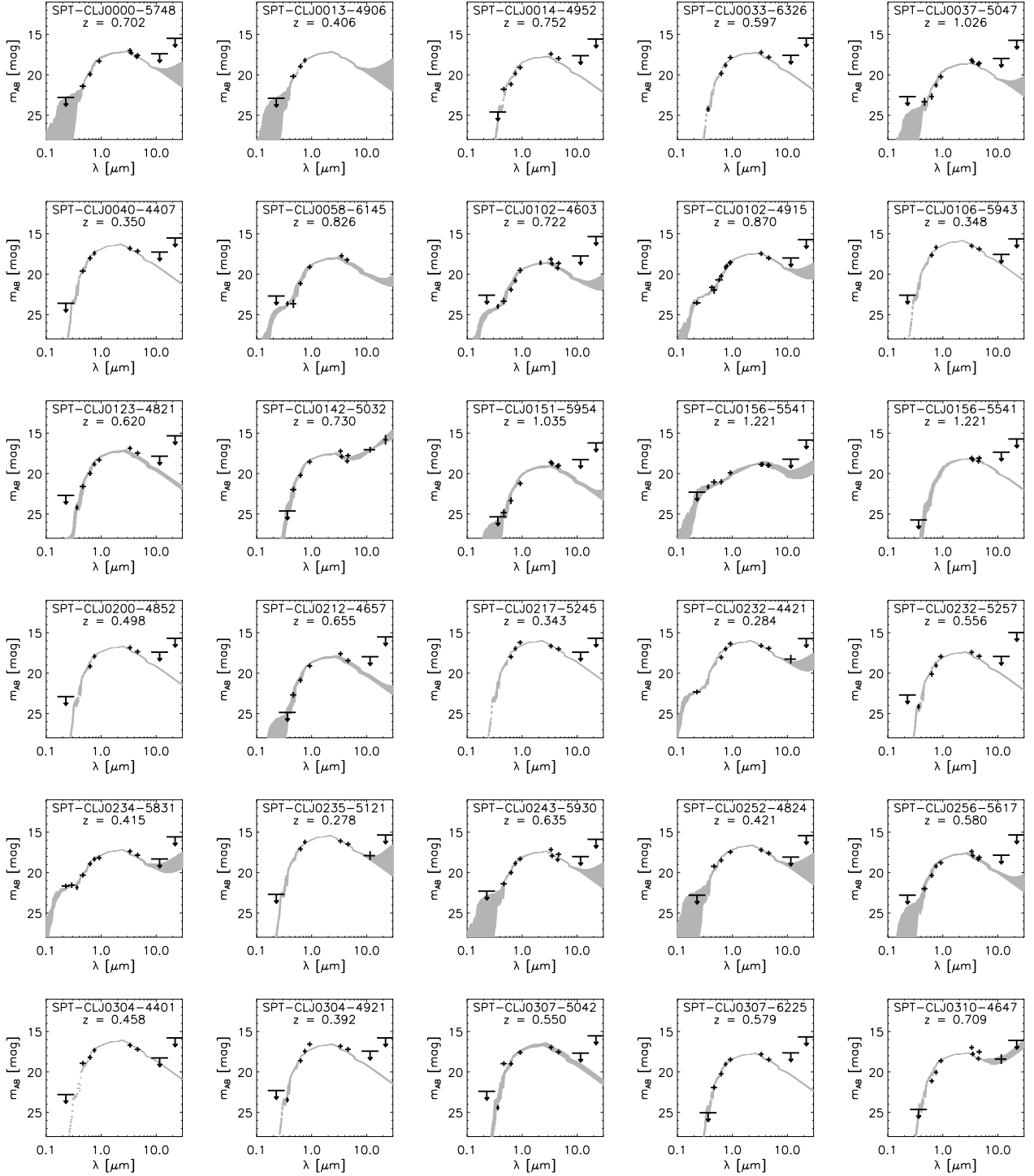}
\caption{Continued next page.}
\label{fig:seds}
\end{figure*}

\renewcommand\thefigure{\thesection.\arabic{figure}}    
\setcounter{figure}{0}    

\begin{figure*}[h!]
\centering
\includegraphics[width=0.99\textwidth,trim=0cm 3.5cm 0cm 0cm,clip=true]{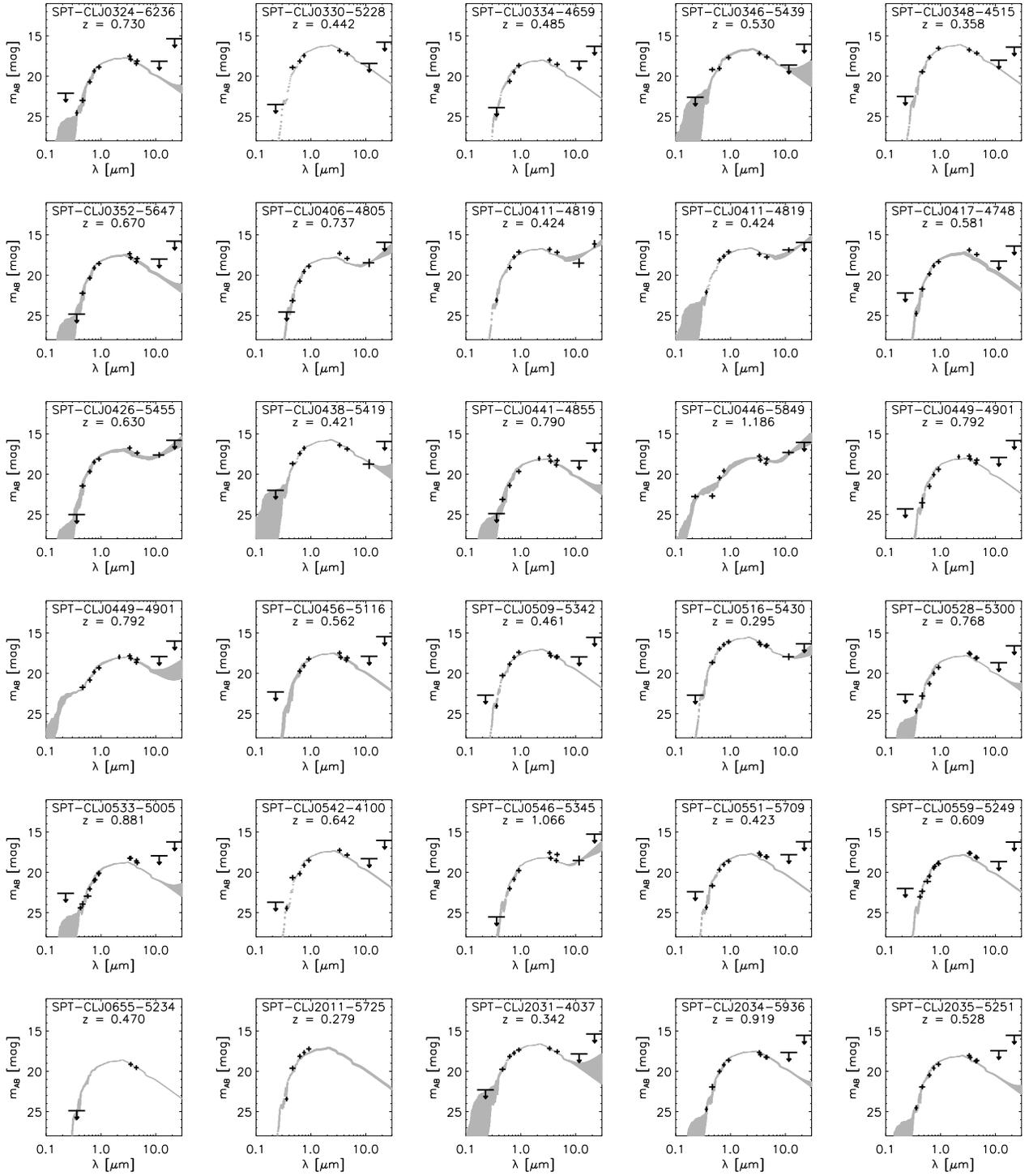}
\caption{Continued next page.}
\label{fig:seds}
\end{figure*}

\renewcommand\thefigure{\thesection.\arabic{figure}}    
\setcounter{figure}{0}    

\begin{figure*}[h!]
\centering
\includegraphics[width=0.99\textwidth,trim=0cm 3.5cm 0cm 0cm,clip=true]{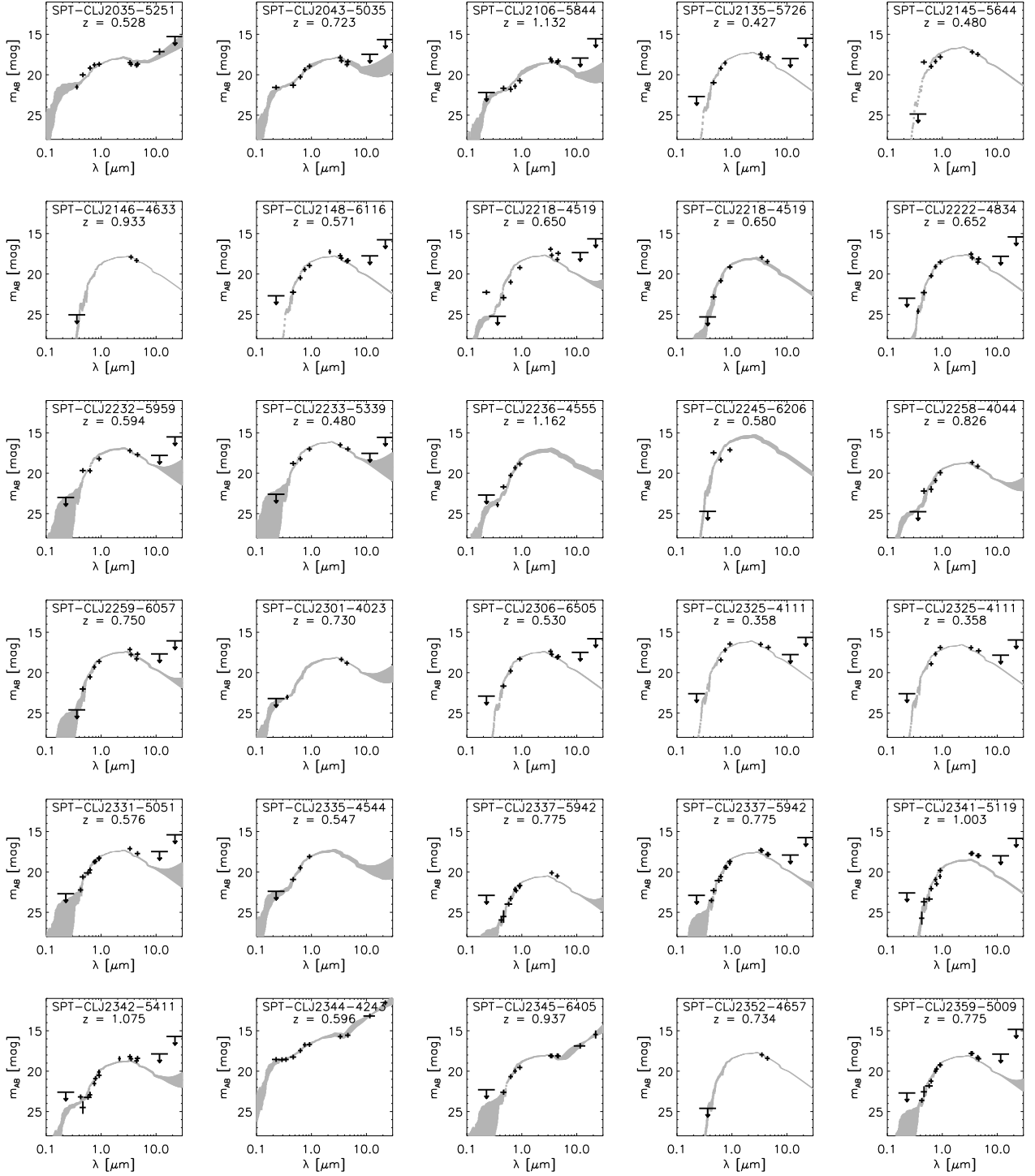}
\caption{Broadband spectral energy distribution for 90 BCGs in this sample. Wavelengths here are in the observed frame. We overplot the best-fitting models from 100 fits, while varying unknown quantities such as the intrinsic reddening ($E(B-V) = 0.3 \pm 0.1$) the mid-IR slope ($\alpha = 2.0 \pm 0.5$) and the formation epoch of the old population ($z_f = 2-5$). The model consists of an old stellar population, formed at $z=z_f$ \citep{leitherer99}, a young stellar population which has been continuously forming stars for 30 Myr \citep{leitherer99}, and a dust component parametrized by a powerlaw in the mid-IR, following \cite{casey12}. }
%\label{fig:seds}
\label{fig:seds}
\end{figure*}

\begin{deluxetable*}{c c c c c c c c}[htb]
\tablecaption{BCG Star Formation Rates and Stellar Masses}
\tablehead{
\colhead{Cluster} &
\colhead{$z_{cluster}$} & 
\colhead{$\alpha_{BCG}$} &
\colhead{$\delta_{BCG}$} & 
\colhead{M$_*$} &
\colhead{SFR$_{UV}$} &
\colhead{SFR$_{[O II]}$} &
\colhead{SFR$_{24\mu m}$}
\\
\colhead{ } &
\colhead{ } &
\colhead{[$^{\circ}$]} & 
\colhead{[$^{\circ}$]} & 
\colhead{[10$^{12}$ M$_{\odot}$]} & 
\colhead{[M$_{\odot}$ yr$^{-1}$]} & 
\colhead{[M$_{\odot}$ yr$^{-1}$]} & 
\colhead{[M$_{\odot}$ yr$^{-1}$]} 
}
\startdata
SPT-CLJ0000-5748 & 0.702 &   0.2501 & $-$57.8093 & 2.54$_{-0.23}^{+0.17}$ & 52$_{-39}^{+89}$ & 51$_{-20}^{+28}$ & $<$85\\
SPT-CLJ0013-4906 & 0.406 &   3.3306 & $-$49.1099 & 0.98$_{-0.10}^{+0.12}$ & $<$50 & $<$8.6 & --\\
SPT-CLJ0014-4952 & 0.752 &   3.7041 & $-$49.8851 & 1.68$_{-0.29}^{+0.10}$ & $<$10. & -- & $<$92\\
SPT-CLJ0033-6326 & 0.597 &   8.4710 & $-$63.4449 & 2.45$_{-0.18}^{+0.09}$ & $<$1300 & $<$12 & $<$43\\
SPT-CLJ0037-5047 & 1.026 &   9.4478 & $-$50.7890 & 1.00$_{-0.18}^{+0.12}$ & 21$_{-15}^{+40}$ & $<$14 & $<$170\\
SPT-CLJ0040-4407 & 0.350 &  10.2080 & $-$44.1307 & 1.79$_{-0.11}^{+0.26}$ & $<$8.5 & -- & $<$14\\
SPT-CLJ0058-6145 & 0.826 &  14.5842 & $-$61.7669 & 1.54$_{-0.47}^{+0.49}$ & 24$_{-13}^{+34}$ & -- & --\\
SPT-CLJ0102-4603 & 0.722 &  15.6779 & $-$46.0710 & 0.73$_{-0.16}^{+0.22}$ & 15$_{-9.0}^{+20}$ & -- & $<$79\\
SPT-CLJ0102-4915 & 0.870 &  15.7407 & $-$49.2720 & 2.43$_{-0.39}^{+0.24}$ & 93$_{-55}^{+120}$ & -- & $<$88\\
SPT-CLJ0106-5943 & 0.348 &  16.6197 & $-$59.7201 & 3.38$_{-0.23}^{+0.18}$ & $<$20 & $<$5.1 & $<$7.3\\
SPT-CLJ0123-4821 & 0.620 &  20.7956 & $-$48.3563 & 2.09$_{-0.64}^{+0.75}$ & $<$3.6 & -- & $<$39\\
SPT-CLJ0142-5032 & 0.730 &  25.5401 & $-$50.5410 & 1.39$_{-0.18}^{+0.48}$ & $<$7.7 & -- & 92$_{-34}^{+54}$\\
SPT-CLJ0151-5954 & 1.035 &  27.8634 & $-$59.9062 & 0.63$_{-0.11}^{+0.28}$ & $<$6.4 & -- & $<$140\\
SPT-CLJ0156-5541 & 1.221 &  29.0436 & $-$55.6985 & 0.33$_{-0.29}^{+0.34}$ & 400$_{-220}^{+410}$ & -- & $<$210\\
SPT-CLJ0156-5541 & 1.221 &  29.0382 & $-$55.7029 & 1.79$_{-0.45}^{+0.25}$ & $<$9.0 & -- & $<$530\\
SPT-CLJ0200-4852 & 0.498 &  30.1421 & $-$48.8712 & 3.12$_{-0.22}^{+0.03}$ & $<$34 & $<$26 & $<$29\\
SPT-CLJ0212-4657 & 0.655 &  33.0986 & $-$46.9537 & 1.47$_{-0.46}^{+0.09}$ & $<$1.0 & -- & $<$49\\
SPT-CLJ0217-5245 & 0.343 &  34.3122 & $-$52.7604 & 2.88$_{-0.07}^{+0.17}$ & $<$450 & -- & $<$9.2\\
SPT-CLJ0232-4421 & 0.284 &  38.0773 & $-$44.3467 & 2.07$_{-0.16}^{+0.12}$ & 17$_{-10.0}^{+19}$ & -- & $<$11\\
SPT-CLJ0232-5257 & 0.556 &  38.2058 & $-$52.9531 & 1.85$_{-0.07}^{+0.11}$ & $<$1.1 & -- & $<$26\\
SPT-CLJ0234-5831 & 0.415 &  38.6761 & $-$58.5236 & 0.94$_{-0.11}^{+0.09}$ & 50$_{-28}^{+71}$ & 59$_{-22}^{+35}$ & $<$7.0\\
SPT-CLJ0235-5121 & 0.278 &  38.9387 & $-$51.3512 & 3.45$_{-0.33}^{+0.03}$ & $<$13 & -- & $<$14\\
SPT-CLJ0243-5930 & 0.635 &  40.8628 & $-$59.5172 & 1.67$_{-0.14}^{+0.09}$ & $<$100 & $<$10. & $<$35\\
SPT-CLJ0252-4824 & 0.421 &  43.2083 & $-$48.4162 & 1.24$_{-0.10}^{+0.14}$ & $<$59 & $<$26 & $<$5.9\\
SPT-CLJ0256-5617 & 0.580 &  44.1056 & $-$56.2978 & 1.45$_{-0.50}^{+0.26}$ & $<$11 & -- & $<$35\\
SPT-CLJ0304-4401 & 0.458 &  46.0707 & $-$44.0256 & 3.22$_{-0.04}^{+0.27}$ & $<$30 & $<$17 & $<$6.1\\
SPT-CLJ0304-4921 & 0.392 &  46.0673 & $-$49.3571 & 2.14$_{-0.17}^{+0.27}$ & $<$35 & -- & $<$15\\
SPT-CLJ0307-5042 & 0.550 &  46.9605 & $-$50.7012 & 1.97$_{-0.15}^{+2.34}$ & $<$64 & -- & $<$25\\
SPT-CLJ0307-6225 & 0.579 &  46.8195 & $-$62.4465 & 1.12$_{-0.08}^{+0.08}$ & $<$4.0 & $<$14 & $<$44\\
SPT-CLJ0310-4647 & 0.709 &  47.6354 & $-$46.7856 & 2.19$_{-0.21}^{+0.08}$ & $<$8.3 & $<$11 & 34$_{-13}^{+21}$\\
SPT-CLJ0324-6236 & 0.730 &  51.0511 & $-$62.5988 & 1.73$_{-0.21}^{+0.29}$ & $<$13 & -- & $<$57\\
SPT-CLJ0330-5228 & 0.442 &  52.7374 & $-$52.4704 & 2.99$_{-0.23}^{+0.13}$ & $<$15 & -- & $<$4.8\\
SPT-CLJ0334-4659 & 0.485 &  53.5457 & $-$46.9958 & 0.93$_{-0.07}^{+0.02}$ & $<$8.7 & 79$_{-30}^{+45}$ & $<$18\\
SPT-CLJ0346-5439 & 0.530 &  56.7308 & $-$54.6487 & 2.41$_{-0.22}^{+0.28}$ & $<$120 & -- & $<$5.7\\
SPT-CLJ0348-4515 & 0.358 &  57.0719 & $-$45.2498 & 2.32$_{-0.17}^{+0.00}$ & $<$22 & $<$4.7 & $<$5.7\\
SPT-CLJ0352-5647 & 0.670 &  58.2397 & $-$56.7977 & 1.93$_{-0.38}^{+0.21}$ & $<$6.8 & -- & $<$50\\
SPT-CLJ0406-4805 & 0.737 &  61.7302 & $-$48.0826 & 1.57$_{-0.24}^{+0.10}$ & $<$10.0 & $<$21 & 41$_{-16}^{+26}$\\
SPT-CLJ0411-4819 & 0.424 &  62.7957 & $-$48.3277 & 2.17$_{-0.14}^{+0.16}$ & -- & $<$5.5 & 15$_{-4.4}^{+4.4}$\\
SPT-CLJ0411-4819 & 0.424 &  62.8154 & $-$48.3175 & 2.54$_{-0.21}^{+0.07}$ & $<$12 & -- & 27$_{-11}^{+5.0}$\\
SPT-CLJ0417-4748 & 0.581 &  64.3461 & $-$47.8132 & 2.33$_{-0.63}^{+0.23}$ & $<$93 & $<$3.8 & $<$28\\
SPT-CLJ0426-5455 & 0.630 &  66.5171 & $-$54.9253 & 2.89$_{-1.20}^{+0.67}$ & $<$4.7 & -- & 51$_{-22}^{+41}$\\
SPT-CLJ0438-5419 & 0.421 &  69.5734 & $-$54.3224 & 3.98$_{-0.20}^{+0.25}$ & $<$31 & $<$23 & $<$24\\
SPT-CLJ0441-4855 & 0.790 &  70.4497 & $-$48.9233 & 1.38$_{-0.23}^{+0.17}$ & $<$13 & -- & $<$61\\
SPT-CLJ0446-5849 & 1.186 &  71.5157 & $-$58.8304 & 1.46$_{-0.59}^{+0.39}$ & 200$_{-140}^{+450}$ & -- & 330$_{-170}^{+270}$\\
SPT-CLJ0449-4901 & 0.792 &  72.2819 & $-$49.0214 & 1.36$_{-0.19}^{+0.04}$ & $<$10. & -- & $<$84\\
SPT-CLJ0449-4901 & 0.792 &  72.2669 & $-$49.0276 & 1.21$_{-0.22}^{+0.14}$ & 88$_{-47}^{+110}$ & $<$27 & $<$81\\
SPT-CLJ0456-5116 & 0.562 &  74.1171 & $-$51.2764 & 1.48$_{-0.34}^{+0.13}$ & $<$79 & $<$6.3 & $<$32\\
SPT-CLJ0509-5342 & 0.461 &  77.3393 & $-$53.7035 & 1.54$_{-0.16}^{+0.11}$ & $<$32 & 33$_{-12}^{+20}$ & $<$16\\
SPT-CLJ0516-5430 & 0.295 &  79.1556 & $-$54.5004 & 2.90$_{-0.17}^{+0.18}$ & $<$13 & $<$27 & 2.5$_{-0.82}^{+1.3}$\\
SPT-CLJ0528-5300 & 0.768 &  82.0222 & $-$52.9981 & 1.59$_{-0.13}^{+0.05}$ & $<$10. & $<$15 & $<$34\\
SPT-CLJ0533-5005 & 0.881 &  83.4033 & $-$50.0958 & 0.75$_{-0.08}^{+0.03}$ & $<$16 & 50$_{-19}^{+29}$ & $<$130\\
SPT-CLJ0542-4100 & 0.642 &  85.7085 & $-$41.0001 & 1.86$_{-0.16}^{+0.11}$ & $<$1.4 & -- & $<$28\\
SPT-CLJ0546-5345 & 1.066 &  86.6573 & $-$53.7588 & 1.70$_{-0.30}^{+0.09}$ & $<$8.0 & -- & 110$_{-53}^{+89}$\\
SPT-CLJ0551-5709 & 0.423 &  87.8931 & $-$57.1451 & 0.69$_{-0.02}^{+0.06}$ & $<$37 & -- & $<$17\\
SPT-CLJ0559-5249 & 0.609 &  89.9301 & $-$52.8242 & 1.11$_{-0.11}^{+0.10}$ & $<$11 & $<$16 & $<$19\\
SPT-CLJ0655-5234 & 0.470 & 103.9760 & $-$52.5674 & 0.36$_{-0.02}^{+0.00}$ & $<$3.1 & -- & --\\
\enddata
%\tablecomments{}
\label{table:sample}
\end{deluxetable*}

\renewcommand\thetable{\thesection.\arabic{table}}    
\setcounter{table}{0}   

\begin{deluxetable*}{c c c c c c c c}[htb]
\tablecaption{BCG Star Formation Rates and Stellar Masses (continued)}
\tablehead{
\colhead{Cluster} &
\colhead{$z_{cluster}$} & 
\colhead{$\alpha_{BCG}$} &
\colhead{$\delta_{BCG}$} & 
\colhead{M$_*$} &
\colhead{SFR$_{UV}$} &
\colhead{SFR$_{[O II]}$} &
\colhead{SFR$_{24\mu m}$}
\\
\colhead{ } &
\colhead{ } &
\colhead{[$^{\circ}$]} & 
\colhead{[$^{\circ}$]} & 
\colhead{[10$^{12}$ M$_{\odot}$]} & 
\colhead{[M$_{\odot}$ yr$^{-1}$]} & 
\colhead{[M$_{\odot}$ yr$^{-1}$]} & 
\colhead{[M$_{\odot}$ yr$^{-1}$]} 
}
\startdata
SPT-CLJ2011-5725 & 0.279 & 302.8620 & $-$57.4196 & 0.71$_{-0.10}^{+0.02}$ & -- & -- & --\\
SPT-CLJ2031-4037 & 0.342 & 307.9720 & $-$40.6252 & 1.33$_{-0.15}^{+0.02}$ & $<$51 & -- & $<$7.5\\
SPT-CLJ2034-5936 & 0.919 & 308.5390 & $-$59.6042 & 1.97$_{-0.30}^{+0.18}$ & $<$4.5 & -- & $<$99\\
SPT-CLJ2035-5251 & 0.528 & 308.7950 & $-$52.8564 & 0.72$_{-0.01}^{+0.05}$ & $<$2.8 & $<$2.4 & $<$41\\
SPT-CLJ2035-5251 & 0.528 & 308.7930 & $-$52.8539 & 0.55$_{-0.05}^{+0.08}$ & 120$_{-60}^{+78}$ & -- & 31$_{-16}^{+29}$\\
SPT-CLJ2043-5035 & 0.723 & 310.8230 & $-$50.5923 & 0.88$_{-0.37}^{+0.26}$ & 210$_{-130}^{+240}$ & 110$_{-47}^{+54}$ & $<$89\\
SPT-CLJ2106-5844 & 1.132 & 316.5190 & $-$58.7411 & 0.83$_{-0.27}^{+0.28}$ & 200$_{-110}^{+240}$ & -- & $<$260\\
SPT-CLJ2135-5726 & 0.427 & 323.9060 & $-$57.4418 & 1.07$_{-0.03}^{+0.08}$ & $<$13 & $<$8.5 & $<$13\\
SPT-CLJ2145-5644 & 0.480 & 326.4660 & $-$56.7482 & 2.30$_{-0.13}^{+0.05}$ & $<$3.6 & $<$26 & --\\
SPT-CLJ2146-4633 & 0.933 & 326.6470 & $-$46.5505 & 1.96$_{-0.25}^{+0.04}$ & $<$9.6 & $<$34 & --\\
SPT-CLJ2148-6116 & 0.571 & 327.1780 & $-$61.2795 & 1.04$_{-0.05}^{+0.07}$ & $<$7.5 & -- & $<$39\\
SPT-CLJ2218-4519 & 0.650 & 334.7470 & $-$45.3145 & 1.12$_{-0.09}^{+0.03}$ & 4.4$_{-2.4}^{+5.0}$ & -- & $<$55\\
SPT-CLJ2218-4519 & 0.650 & 334.7500 & $-$45.3162 & 1.06$_{-0.27}^{+0.25}$ & $<$1.0 & -- & --\\
SPT-CLJ2222-4834 & 0.652 & 335.7110 & $-$48.5764 & 1.47$_{-0.08}^{+0.05}$ & $<$3.1 & -- & $<$52\\
SPT-CLJ2232-5959 & 0.594 & 338.1410 & $-$59.9980 & 2.27$_{-0.26}^{+0.38}$ & $<$110 & $<$18 & $<$33\\
SPT-CLJ2233-5339 & 0.480 & 338.3150 & $-$53.6526 & 2.58$_{-0.20}^{+0.27}$ & $<$86 & $<$12 & $<$13\\
SPT-CLJ2236-4555 & 1.162 & 339.2140 & $-$45.9295 & 3.29$_{-0.97}^{+0.64}$ & 50$_{-30}^{+69}$ & -- & --\\
SPT-CLJ2245-6206 & 0.580 & 341.2590 & $-$62.1272 & 7.09$_{-0.86}^{+4.64}$ & $<$5.2 & -- & --\\
SPT-CLJ2258-4044 & 0.826 & 344.7010 & $-$40.7418 & 0.62$_{-0.07}^{+0.08}$ & 10.$_{-5.7}^{+12}$ & -- & --\\
SPT-CLJ2259-6057 & 0.750 & 344.7540 & $-$60.9595 & 2.15$_{-0.25}^{+0.44}$ & $<$11 & -- & $<$84\\
SPT-CLJ2301-4023 & 0.730 & 345.4700 & $-$40.3868 & 1.04$_{-0.15}^{+0.07}$ & 36$_{-19}^{+53}$ & -- & --\\
SPT-CLJ2306-6505 & 0.530 & 346.7230 & $-$65.0882 & 1.34$_{-0.05}^{+0.06}$ & $<$14 & $<$2.6 & $<$39\\
SPT-CLJ2325-4111 & 0.358 & 351.2990 & $-$41.2037 & 3.01$_{-0.13}^{+0.10}$ & $<$9.8 & $<$11 & $<$6.7\\
SPT-CLJ2325-4111 & 0.358 & 351.3000 & $-$41.1991 & 1.96$_{-0.03}^{+0.26}$ & $<$14 & $<$5.0 & $<$8.4\\
SPT-CLJ2331-5051 & 0.576 & 352.9630 & $-$50.8650 & 1.53$_{-0.08}^{+0.06}$ & 23$_{-14}^{+36}$ & -- & $<$46\\
SPT-CLJ2335-4544 & 0.547 & 353.7850 & $-$45.7391 & 1.40$_{-0.37}^{+0.22}$ & 41$_{-27}^{+61}$ & $<$6.7 & --\\
SPT-CLJ2337-5942 & 0.775 & 354.3550 & $-$59.7058 & 0.13$_{-0.01}^{+0.01}$ & 1.00$_{-0.66}^{+1.5}$ & -- & --\\
SPT-CLJ2337-5942 & 0.775 & 354.3650 & $-$59.7013 & 2.08$_{-0.28}^{+0.09}$ & $<$4.9 & $<$40 & $<$76\\
SPT-CLJ2341-5119 & 1.003 & 355.3010 & $-$51.3291 & 1.15$_{-0.25}^{+0.10}$ & $<$2.5 & -- & $<$170\\
SPT-CLJ2342-5411 & 1.075 & 355.6910 & $-$54.1847 & 0.90$_{-0.22}^{+0.13}$ & 33$_{-19}^{+42}$ & $<$17 & $<$240\\
SPT-CLJ2344-4243 & 0.596 & 356.1830 & $-$42.7201 & 3.95$_{-2.43}^{+1.85}$ & 2000$_{-1200}^{+2400}$ & 1700$_{-710}^{+1100}$ & 2000$_{-730}^{+860}$\\
SPT-CLJ2345-6405 & 0.937 & 356.2510 & $-$64.0927 & 1.20$_{-0.16}^{+0.11}$ & $<$71 & -- & 250$_{-89}^{+130}$\\
SPT-CLJ2352-4657 & 0.734 & 358.0680 & $-$46.9602 & 1.64$_{-0.14}^{+0.07}$ & $<$9.5 & -- & --\\
SPT-CLJ2359-5009 & 0.775 & 359.9280 & $-$50.1672 & 1.25$_{-0.15}^{+0.05}$ & 8.2$_{-6.0}^{+13}$ & $<$21 & $<$84\\
\enddata
\tablecomments{Positions and star formation rates of BCGs used in this study. UV- and [O\,\textsc{ii}]-derived SFRs are corrected for intrinsic extinction assuming $E(B-V) = 0.3 \pm 0.1$ and a gray extinction curve from \cite{calzetti00}.
Systematic uncertainties, which are dominated by the extinction correction in the blue bands and extrapolation to 24$\mu$m in the IR band, are quoted for all detections. These uncertainties are discussed in detail in \S2.5.}
\label{table:sample}
\end{deluxetable*}

\end{appendices}
\end{document}